\documentclass[sensors,article,accept,moreauthors,pdftex]{Definitions/mdpi} 

\usepackage{hyperref}
\usepackage{siunitx}

\firstpage{1} 
\pubvolume{1}
\issuenum{1}
\articlenumber{0}
\pubyear{2021}
\copyrightyear{2021}
\externaleditor{{Academic Editor: Marco Leo} 
}
\datereceived{19 March 2021} 
\dateaccepted{28 April 2021} 
\datepublished{} 
\hreflink{https://doi.org/}

\Title{SpeakingFaces: A Large-Scale Multimodal Dataset of Voice Commands with Visual and Thermal Video Streams}

\TitleCitation{SpeakingFaces: A Large-Scale Multimodal Dataset of Voice Commands with Visual and Thermal Video Streams}

\Author{Madina Abdrakhmanova$^{1}$, Askat Kuzdeuov $^{1}$\orcidB{}, Sheikh Jarju $^{1}$, Yerbolat Khassanov $^{1}$\orcidD{}, Michael Lewis $^{2}$ and~Huseyin~Atakan~Varol $^{1,2,}$*}

\AuthorNames{Madina Abdrakhmanova, Askat Kuzdeuov, Sheikh Jarju, Yerbolat Khassanov, Michael Lewis and~Huseyin~Atakan~Varol}

\AuthorCitation{Abdrakhmanova, M.; Kuzdeuov, A.; Jarju, S.; Khassanov Y.; Lewis M.; Varol H.A.}

\address{%
$^{1}$ \quad Institute of Smart Systems and Artificial Intelligence, Nazarbayev University, Nur-Sultan 010000, Kazakhstan; madina.abdrakhmanova@nu.edu.kz (M.A.); askat.kuzdeuov@nu.edu.kz (A.K.); sheikh.jarju@nu.edu.kz (S.J.); yerbolat.khassanov@nu.edu.kz (Y.K.)\\
$^{2}$ \quad School of Engineering and Digital Sciences, Nazarbayev University, Nur-Sultan 010000, Kazakhstan; mlewis@nu.edu.kz}

\corres{Correspondence: ahvarol@nu.edu.kz}

\abstract{
We present SpeakingFaces as a publicly-available large-scale multimodal dataset developed to support machine learning research in contexts that utilize a combination of thermal, visual, and audio data streams; examples include human--computer interaction, biometric authentication, recognition systems, domain transfer, and speech recognition. 
SpeakingFaces is comprised of aligned high-resolution thermal and visual spectra image streams of fully-framed faces synchronized with audio recordings of each subject speaking approximately 100 imperative phrases.
Data were collected from 142 subjects, yielding over 13,000 instances of synchronized data ($\sim$3.8 TB).
For technical validation, we demonstrate two baseline examples.
The first baseline shows classification by gender, utilizing different combinations of the three data streams in both clean and noisy environments.
The second example consists of thermal-to-visual facial image translation, as an instance of domain~transfer.
}


\begin{document}

\section{Introduction}
\label{sec:Introduction}

The fusion of visual, thermal, and audio data sources opens new opportunities for multimodal data use in a wide range of applications, including human--computer interaction (HCI), biometric authentication, and recognition systems. Multimodal systems are inclined to be more robust and reliable, as different streams can provide complementary information, and failures in one stream can be mitigated by others~\cite{chen2020multi}. Recently introduced high-resolution thermal cameras provide a more granular association of temperature values with facial features. It has been demonstrated that the combination of thermal and visual data can overcome the respective drawbacks of each individual stream~\cite{gade2014thermal}. The addition of visual data to speech signals has also been shown to have a positive impact on improving person verification and speech recognition models~\cite{chen2020multi, 8683477,afouras2018deep}. 

Furthermore, with the emergence of virtual assistants, voice search, and voice command control in smart devices and other Internet of Things (IoT) technologies, voice-enabled applications have attracted considerable attention. The combination of visual and thermal facial data with the corresponding voice records could enable a more nuanced analysis of speech in applications such as the dictation of instructions to smart devices in sub-optimal physical environments, resolution of multi-talker overlapping speech (to distinguish individual speakers and respective intentionality), and improving the performance of automated speech recognition~\cite{afouras2018deep,DBLP:journals/tmm/TaoB21}.

With the miniaturization of uncooled thermal imaging chips, companies started equipping smartphones with thermal cameras, thus introducing mobile devices that combine all of the three modalities. A developer in thermal imaging solutions, FLIR, developed the FLIR ONE Pro thermal camera that can be connected to any Android or iOS smartphone~\cite{one_pro}. The construction machinery and equipment company Caterpillar introduced CAT S62 Pro~\cite{s62}, an Android phone with an integrated FLIR Lepton 3.5 professional-grade sensor~\cite{lepton}. Both devices currently support relatively low resolution thermal cameras ($160\times120$), but, given recent trending of the technology, their successors will likely be of higher-resolution, and thereby could support the more data-intensive multimodal applications. To facilitate such research, we introduce SpeakingFaces, a large-scale dataset consisting of spatially aligned thermal and visual image sequences accompanied by voice command recordings. 



To date, there are no large-scale datasets that combine all three data streams, consisting of synchronized visible-spectrum images, thermal images, and audio tracks. Most of the existing visual-thermal facial datasets are constrained by the issues of a small number of subjects, too few unique instances (thus inhibiting data-hungry machine learning algorithms), low resolution of thermal images, little variability in head postures, or a lack of alignment. These datasets are summarized in Table ~\ref{table:thvis}.
\end{paracol}
  \begin{specialtable}[H]
    \tablesize{\footnotesize}
    \widetable
\caption{\label{table:thvis} Publicly available datasets where visual and thermal images were acquired simultaneously.}
\setlength{\cellWidtha}{\columnwidth/7-2\tabcolsep+0.2in}
\setlength{\cellWidthb}{\columnwidth/7-2\tabcolsep-.2in}
\setlength{\cellWidthc}{\columnwidth/7-2\tabcolsep+0.0in}
\setlength{\cellWidthd}{\columnwidth/7-2\tabcolsep+0.4in}
\setlength{\cellWidthe}{\columnwidth/7-2\tabcolsep-.2in}
\setlength{\cellWidthf}{\columnwidth/7-2\tabcolsep-.2in}
\setlength{\cellWidthg}{\columnwidth/7-2\tabcolsep+0.0in}
\scalebox{1}[1]{\begin{tabularx}{\columnwidth}{
>{\PreserveBackslash\raggedright}m{\cellWidtha}
>{\PreserveBackslash\centering}m{\cellWidthb}
>{\PreserveBackslash\centering}m{\cellWidthc}
>{\PreserveBackslash\centering}m{\cellWidthd}
>{\PreserveBackslash\centering}m{\cellWidthe}
>{\PreserveBackslash\centering}m{\cellWidthf}
>{\PreserveBackslash\centering}m{\cellWidthg}}
\toprule
\textbf{Datasets} & \textbf{$\#$ Subjects} & \textbf{$\#$ Image Pairs} & {\textbf{Thermal} \textbf{Resolution}}  & \textbf{$\#$ Poses} & \textbf{$\#$ Trials} & \textbf{Aligned} \\\midrule
Carl~\cite{carl} & 41 & 2460 & $160\times120$ & 1 & 1 & no\\ 
VIS-TH~\cite{Mallat18} & 50 & 2100 & $160\times120$ & 4 & 2 & yes\\
IRIS~\cite{iris} & 30 & 4228 & $320\times240$ & 11 & 1 & no\\
USTC-NVIE~\cite{5523955} & 215 & N/A & $320\times240$ & 1 & 1 & no\\
Tufts~\cite{panetta2018comprehensive} & 100 & 3600 & $336\times256$ & 9 & 1 & no \\
UL-FMTV~\cite{ghiass2018universite} & 238 & N/A & $640\times512$ &1 &$>$1 & N/A\\
ARL-VTF~\cite{poster2021large} & 395 & 549,712 & $640\times512$ & 3 & 1 & no\\
SpeakingFaces &  142 & 4,581,595 & $464\times348$ & 9 & 2 & yes\\
\bottomrule
\end{tabularx}}
\end{specialtable}

\begin{figure}[H]
    \widefigure
    \includegraphics[width=\linewidth,trim={0cm -0.5cm 0cm 0cm},clip=true]{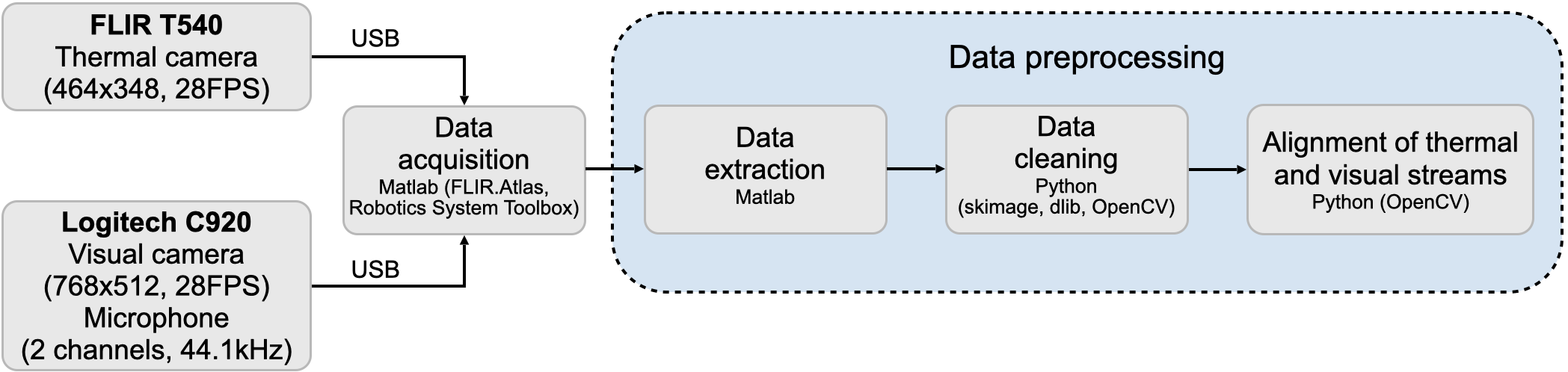}
    \caption{Data pipeline for SpeakingFaces.\label{fig:pipeline}}
\end{figure}

\begin{paracol}{2}
\switchcolumn

The Carl~\cite{carl} and VIS-TH~\cite{Mallat18} databases have the fewest image pairs and the lowest resolution of thermal camera, although the latter involved two trials of each person with four head postures and aligned image pairs. While the IRIS~\cite{iris} dataset has the smallest number of subjects, each subject's face was captured from 11 angles. The USTC-NVIE~\cite{5523955} dataset is comprised of a large number of subjects, but the data were collected using a low-resolution camera from a single position in a single trial. The Tufts~\cite{panetta2018comprehensive} dataset contains a variety of head poses, but a low number of images per subject. UL-FMTV~\cite{ghiass2018universite} involves multiple trials, but only from the frontal position. Although ARL-VTF~\cite{poster2021large} has the largest number of subject and images, as well as the highest thermal resolution, it lacks in the number of head postures and trials. 


Popular audio-visual datasets include Grid~\cite{cooke2006audio}, the Oxford-BBC Lip Reading in the Wild (LRW)~\cite{Chung16} and the Oxford-BBC Lip Reading Sentences (LRS)~\cite{chung2017lip}. The Grid dataset consists of 34 subjects, each uttering 1000 sentences. Each sentence has the same structure: verb (4 types) + color (4 types) + preposition (4 types) + alphabet (25 types) + digit (10~types) + adverb (4 types). The main shortcomings are that data acquisition was conducted in a controlled lab environment, and the utterances are unnatural due to the restricted structure of the sentences.

The LRW dataset has a much greater variety in vocabulary and subjects. It is comprised of over one thousand different speakers and up to 400,000 utterances. However, each utterance is an isolated word, 500 unique instances in total, selected from the BBC television. This constraint was addressed in LRS, a large-scale dataset (100,000 natural sentences and a vocabulary size of around 17,000 words) designed to enable lip reading in an unconstrained natural environment. Neither LRW nor LRS contains thermal data. 

SpeakingFaces is designed to overcome the limitations of the existing multimodal datasets. 
SpeakingFaces consists of 142 subjects, gender-balanced and ethnically diverse. Each subject is recorded in close proximity from nine different angles uttering approximately 100 English phrases or imperative commands, yielding over 13,000 instances of spoken commands, and more than 45 h of video sequences (over 3.7 million image pairs). The spoken phrases are taken from the Stanford University open source digital assistant database~\cite{campagna2017almond}, along with publicly available command sets for the Siri virtual assistant~\cite{siri1,siri2}, chosen to reflect the likely use-case of humans interacting with devices. 

The SpeakingFaces dataset can be used in a wide range of multimodal machine learning contexts, especially those related to HCI, biometrics, and recognition systems. The main contributions of this work are summarized below:
\begin{itemize}
    \item We introduce SpeakingFaces, a large-scale publicly available dataset of voice commands accompanied by streams of visible and thermal image sequences.   
    \item We prepare the dataset by aligning the video streams to minimize the pixel-to-pixel alignment errors between the visual and thermal images. This procedure allows for automatic annotation of thermal images using facial bounding boxes extracted from their visual pairs. 
    \item We provide full annotations on each utterance of a command. 
    \item We present two baseline tasks to illustrate the utility and reliability of the dataset: a classifier for gender using all the three data streams, and an instance of thermal-to-visual image translation as an example of domain transfer. The data used for the latter experiment is publicly available and can be used as a benchmark for image translation~models.  

\end{itemize} 

The rest of this paper is organized as follows. Section \ref{sec2} describes the data collection setup and protocol, the data preparation procedure, and the database structure. Section \ref{sec:Preliminary Experiments} presents and discusses the results of the two baseline tasks, as well as the limitations of our work. Section \ref{sec:Conclusion} concludes the paper and discusses future work.

\section{Materials and Methods}\label{sec2}
In this section, we provide details on the data collection setup and protocol, the data preparation procedure, and the database structure.
Figure~\ref{fig:pipeline} presents the data pipeline in our work.
For sessions that involved uttering commands, the preparation of acquired data begins with the extraction of synchronized video-audio segments. All video segments from both sessions are then converted into image sequences. Next, the visual images are aligned with their thermal pairs using heated ArUco markers~\cite{garrido2014automatic}.

\end{paracol}
\nointerlineskip
\begin{paracol}{2}
\switchcolumn
\vspace{-12pt}
\subsection{Data Acquisition}
\label{sec:Data Acquisition}
The project was conducted with the explicit approval of the Institutional Research Ethics Committee of Nazarbayev University. Each subject participated voluntarily and was informed of the data collection and use protocols, including the acquisition of identifiable images which will be shared as a dataset. The informed consent forms were signed by each~subject.

The setup for the data collection process is shown in Figure~\ref{fig:setup}. Subjects were seated in front of the data collection setup at a distance of approximately one meter. The room temperature was regulated at $\SI{25}{\celsius}$. A subject was illuminated by the ceiling lights in the laboratory room. To ensure the same illumination conditions for all recording sessions, the location and intensity of the light source were fixed. The setup consisted of a metal-framed grid to facilitate camera orientation and two 85$'$$'$ video screens upon which textual phrases were simultaneously presented; two screens were used to minimize the need for subjects to turn their heads while reading the phrases.

\begin{figure}[H]
    \includegraphics[width=.95\linewidth]{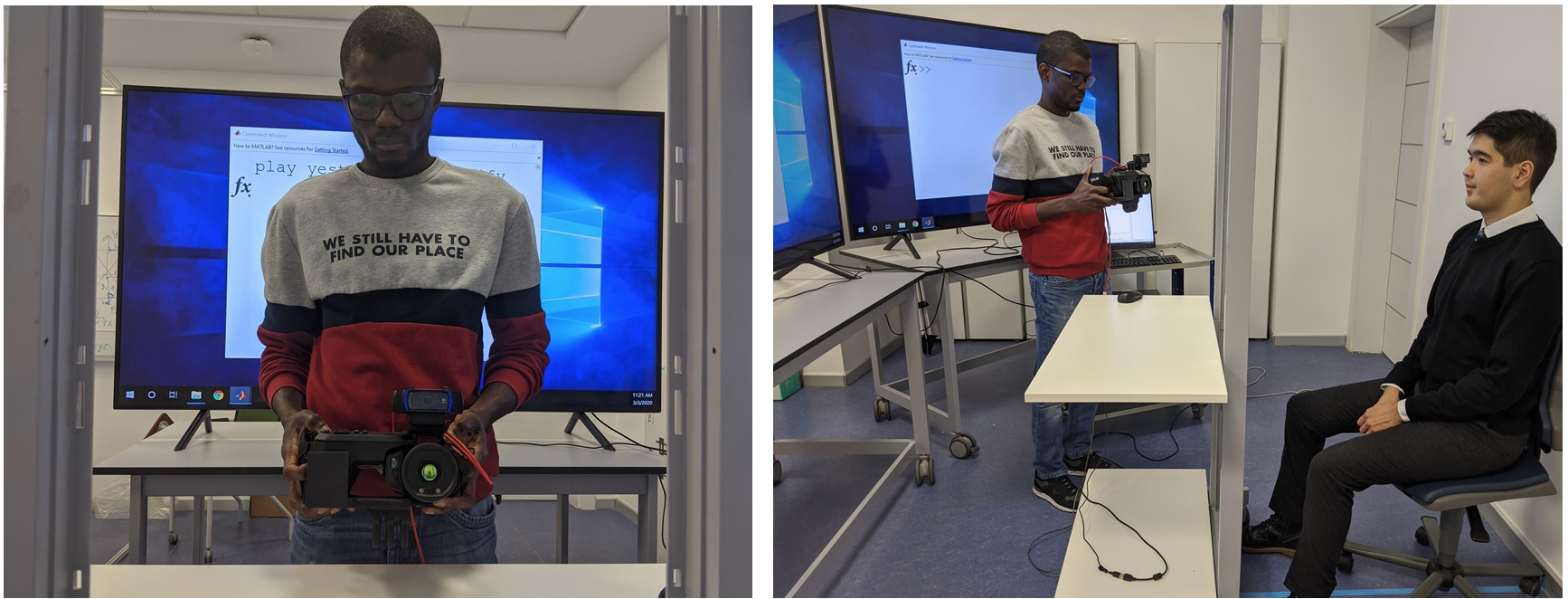}
    \caption{Data acquisition setup for SpeakingFaces.}
    \label{fig:setup}
\end{figure}

The video setup consisted of a FLIR T540 thermal camera (resolution $464\times348$, wave band 7.5--14 $\upmu \text{m}$, and 24$^{\circ}$ field of view) with an attached visual spectrum camera, a Logitech C920 Pro HD web-camera (resolution $1920\times1080$ and field of view 78$^{\circ}$), which has a built-in dual stereo microphone (44.1 kHz). The web-camera was attached on top of the thermal camera to facilitate the subsequent alignment of the image pairs. The original resolution of the web-camera was decreased to $768\times512$ in order to maximize and align the frame rates for both cameras, while preserving the region-of-interest (RoI)--that is, the face. 
The synchronization of the three data streams was achieved using the Robotics System Toolbox of MATLAB~\cite{ros}. The data acquisition code began by launching an audio recorder and then proceeded with iterative an capture of images using both cameras, at a fixed frequency of 28 frames per second (fps). Once the calculated number of frames was captured, the audio recorder stopped. The source code for data acquisition is provided in our GitHub repository 
 (\url{https://github.com/IS2AI/SpeakingFaces}, accessed August 9, 2020). 

The camera operator proceeded manually through a series of nine positions to cover a face from all major angles (similar to Panetta et al.~\cite{panetta2018comprehensive}), as shown in Figure~\ref{fig:3Dfig}. 
The duration of data collection for each position was set to 900 frames. Given the data collection rate of 28 fps for both cameras, this is equivalent to approximately 32 s of video, yielding on average 4.5 min of total video per subject. The subject sat on a chair as shown in Figure~\ref{fig:setup}. The height of the chair was adjusted in order to position the top of the subject's head at a predefined mark. 

It was important to capture the whole face from each of the nine angles. Due to variability in size among the participants, a manual collection process was consciously chosen over the use of fixed positions (such as tripods or mounting frames), or the use of a motorized system covering pre-determined angles. The operator oriented the side, top, and bottom shots to ensure that all of the facial landmarks were fully framed. As a result, there is slight variation of the nine angles, from subject to subject, due to the adjustment of the orientation and framing. Figure~\ref{fig:9pos} presents the image pairs from nine predefined positions of nine subjects.
\begin{figure}[H]
    \includegraphics[width=.95\linewidth]{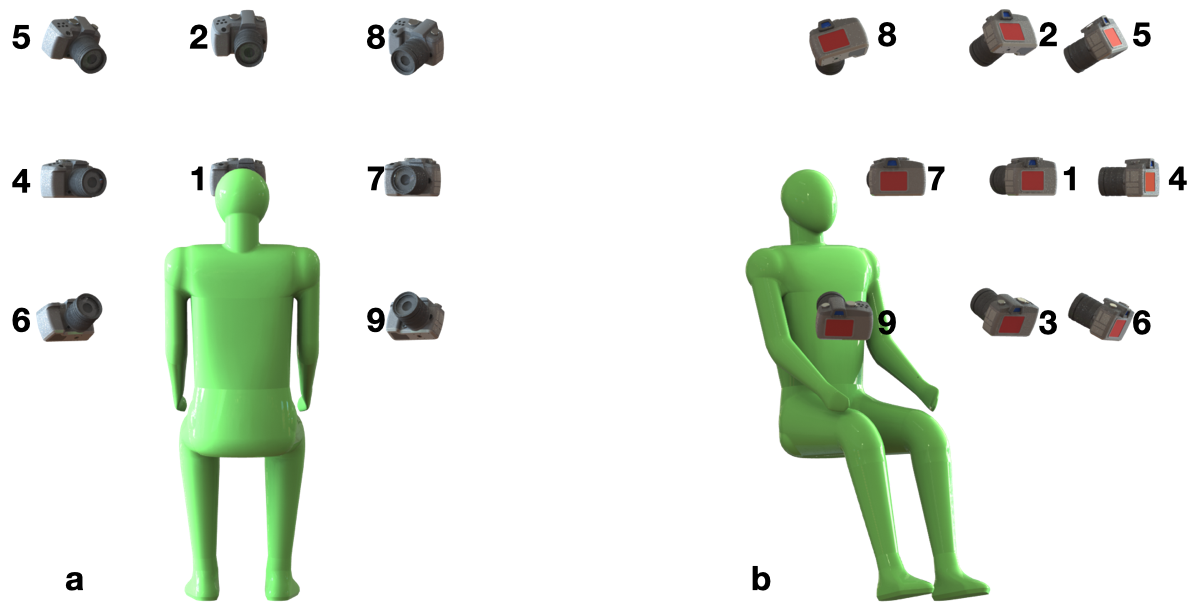}
    \caption{The 3D diagram of the nine camera positions with respect to a subject. (\textbf{a}) Back view. (\textbf{b})~Side view.\label{fig:3Dfig}}
\end{figure}

\vspace{-6pt}
\begin{figure}[H]
    \includegraphics[width=\linewidth]{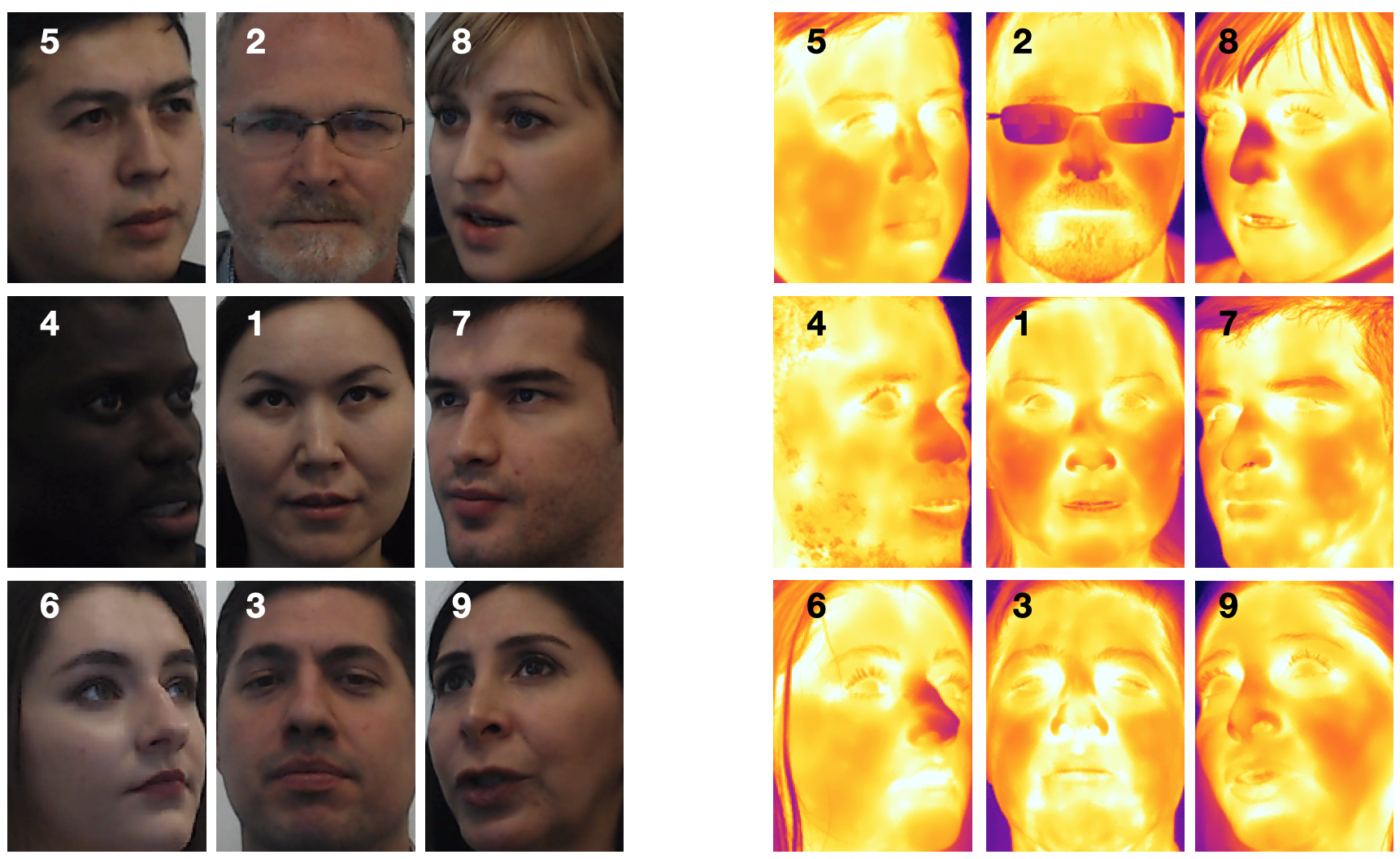}
    \caption{The pairs of visual and thermal facial images of nine subjects taken from the predefined nine positions.\label{fig:9pos}}
\end{figure}

Each subject participated in two types of sessions during a single trial. In the first session, subjects were asked to remain silent and still, with the operator capturing visual and thermal video streams through the procession of nine collection angles. The second session consisted of the subject reading a series of commands presented one-by-one on the video screens, while the visual, thermal, and audio data were being collected from the same nine camera positions. 

Each subject participated in two trials, conducted on different days, at least two weeks apart, consisting of both types of sessions. This was done in order to account for the day-to-day variations of the subjects. For example, some subjects wore glasses during one session, but not in the other. Some subjects changed their hairstyle in between the sessions. Thus, for each subject, there are two trials with three data streams (audio, visible-spectrum video, and thermal-spectrum video) and two trials with two data streams (visual and~thermal).

The commands were sourced from Thingpedia, an open and crowd-sourced knowledge base for virtual assistants~\cite{campagna2017almond}. Thingpedia is a part of the Almond project at Stanford University, and currently includes natural language interfaces for over 128 devices. The interfaces are comprised of utterances grouped by different command types. We selected those that correspond to action and query commands for each device. This resulted in nearly 1500 unique commands: 1297 of them were set aside for training, while the rest were used for test and validation. The total count for the latter part (test and validation) was increased to 500 by utilizing publicly available commands for Siri~\cite{siri1,siri2}. 
We split them in half, such that the commands from Thingpedia would appear evenly in the test and validation sets. 
The commands in the training, validation, and test sets are unique, that is, they do not overlap.

To ensure that each command is uttered by multiple speakers with varying accent, gender, and ethnicity, it was duplicated eight times, as it had been done for the LRW dataset. 
This approach provided data volumes sufficient for 142 subjects. The resulting list of commands for each set was randomly shuffled and partitioned into small groups as follows. First, the duration of a command was calculated by multiplying the number of characters in the command by the average speed of reading, empirically estimated at 5 frames per character. Then, it was used to fit as many commands as possible within the 900-frame window allocated for each position. To enable the automatic extraction of commands, the starting and ending frames for each command in a group were marked. Figure~\ref{fig:time} shows a sequence of images with 0.5-second intervals illustrating different patterns of the lips during the utterance of a voice command.


\begin{figure}[H]
\includegraphics[width=0.8\linewidth]{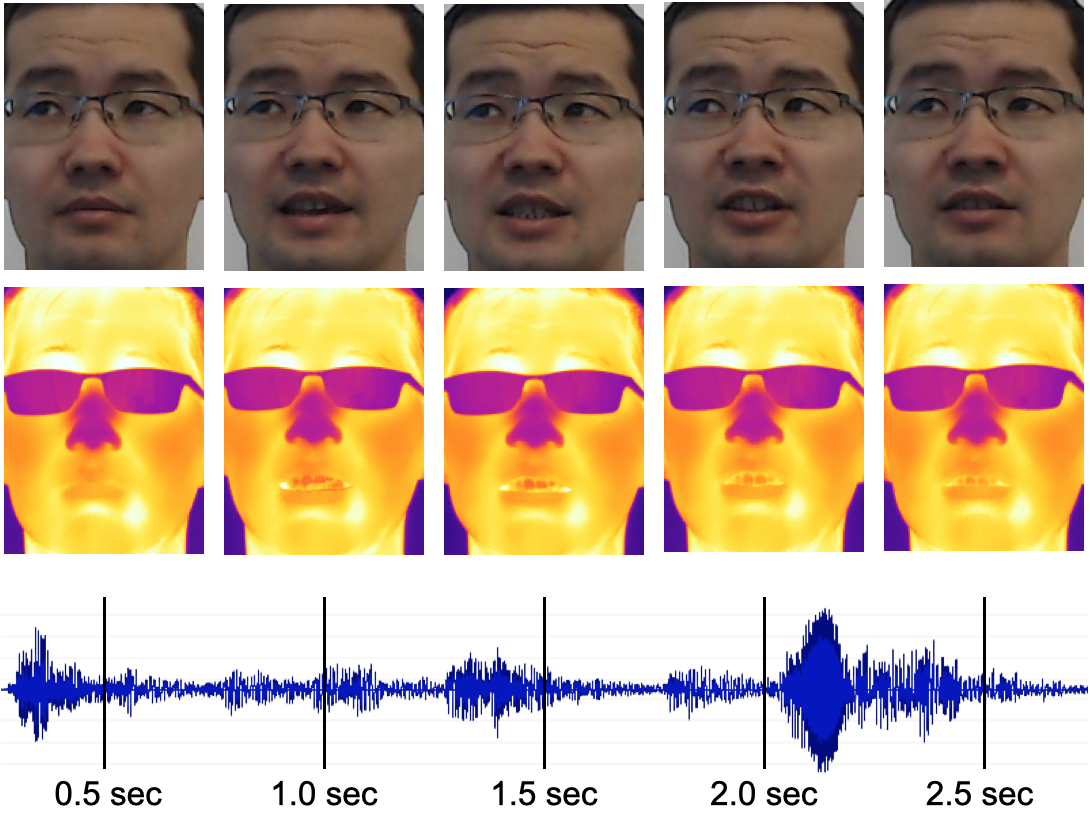}
\caption{Snapshots from the visual and thermal streams with 0.5-second intervals during the utterance of ``stop the kitchen fan from turning''.}
\label{fig:time}
\end{figure}

\subsection{Data Preprocessing} 
\label{sec:Data Preprocessing}



In trials where subjects sat still, without uttering any commands, the raw videos were converted to sequences of images (900 images per position). 
In the speaking trials, the raw video and audio files were first cut into short segments based on the annotations of the start and end frames of each utterance. 
Then, due to the variation in reading speed among our subjects, the audio segments were manually trimmed, with at most one second left at the end of each utterance. The files were also validated to be complete, with minor text noise, such as hesitations or stumbling. The valid recordings were re-transcribed to capture the exact utterance in order to further minimize noise in the text data. The video segments were then converted into image sequences based on the duration of the resulting audio files. If the text noise was substantial, beyond routine hesitations and stumbling, then the utterance was eliminated from the final version of the dataset.

Upon the examination of image frames, we encountered four major artifacts: camera freeze (in thermal), blurriness, flickering, and a slight cut of a chin (in visual). Camera freeze detection in thermal images was based on the analysis of consecutive frames with the Structural Similarity Index of scikit-image~\cite{ssim_skimage, compare_img}. Blur detection was implemented using the variance of the Laplacian method with OpenCV~\cite{blur_opencv}. 
Flickering was detected by keeping track of facial bounding boxes with the dlib library~\cite{dlib09} while processing a sequence of visual frames. A significant shift in the coordinates of a bounding box indicated that the artifact was present, and the corresponding frames were marked. The results showed that flickering happened only at the beginning of a recording, before subjects started speaking. Thus, the affected frames were deleted, and the corresponding audio files were trimmed to safely remove this artifact from the final version of the dataset. The detection of cropped chins was implemented by extracting facial landmarks with the dlib library from visual images, before they were aligned with their thermal pairs. If any coordinates of the landmarks in the chin region were beyond the boundaries of an image, then it meant that this landmark was not present in the image. Overall, each artifact detected by the code was validated by one of the authors of this manuscript. The code for all the artifact detection routines can be found in our GitHub repository 
\linebreak (\url{https://github.com/IS2AI/SpeakingFaces}, accessed August 9, 2020).

Image pairs from the two cameras were aligned using a method involving the estimation of a planar homography~\cite{szeliski2010computer}. This process requires matching at least four paired pixel coordinates that correspond to features present in both thermal and visual images. For visual cameras, a printed image of a chessboard is a common calibration object due to its sharp and distinctive features~\cite{chess_opencv}. However, the crispness of the edges degrades significantly when heated and captured by a thermal camera. One way to overcome this issue is to construct a composite chessboard of two different materials~\cite{gade2014thermal}. Another approach utilizes a board with a fixed pattern of holes~\cite{hwang2015multispectral}; when the board is heated, the features become more apparent to a thermal sensor. 

\textls[-8]{For our collection process, we chose ArUco markers, which are synthetic square markers with a black border and a unique binary (black and white) inner matrix that determines its unique identifier~(ID)~\cite{garrido2014automatic}. These markers have been used for robotics~\cite{kuzdeuov2020neural, BABINEC20141},} autonomous systems~\cite{bacik2017}, and virtual reality~\cite{lupu2017} thanks to their robustness and versatility. Each detected marker provides the ID and pixel coordinates of its four corners. Detecting these markers in both types of images simplifies the process of obtaining paired pixel~coordinates. 

We utilized 12 ArUco markers as shown in Figure~\ref{fig:markers}. In order to detect them in a thermal image, a printed copy of the markers was heated using a flood light (Arrilite 750~Plus) and then captured with the setup consisting of thermal and visual cameras. The thermal image was converted to the grayscale and then negated so that the markers would appear similar to the visual image, with black borders and a correctly colored binary matrix. The ArUco detection algorithm successfully found all the 12 markers in both images and generated $48$ matched pixel coordinate pairs ($12\times4$) in total. These points were fed to OpenCV’s findHomography function~\cite{homography_opencv} to estimate the homography matrix and warpPerspective function~\cite{warp_opencv} to apply a perspective transformation onto a visual image. The source code for collecting and pre-processing data is available in our GitHub repository (\url{https://github.com/IS2AI/SpeakingFaces}, accessed August 9, 2020) under the MIT license.
\begin{figure}[H]
\includegraphics[width=.95\linewidth]{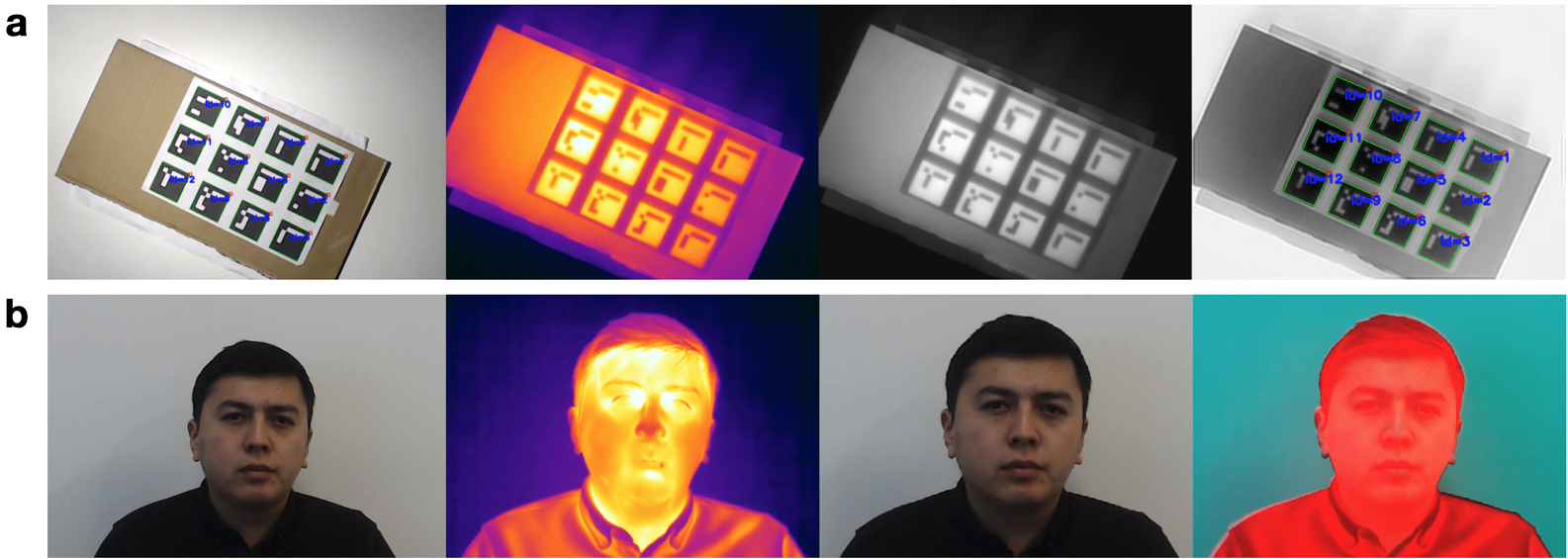}
\caption{(\textbf{a}) Left to right: the visual image with identified ArUco markers; the original thermal image; a grayscale version of the thermal image; and a complement of the grayscale image with identified ArUco markers. (\textbf{b}) Left to right: the original visual image; the original thermal image; the result of aligning the visual image with it’s thermal pair; and the result of replacing the red channel of the visual image with the red channel of the thermal image.}
\label{fig:markers}
\end{figure}


\subsection{Database Structure}
\label{sec:Database Structure}
The SpeakingFaces dataset is available through the server of the Institute of Smart Systems and Artificial Intelligence (ISSAI) under Creative Commons Attribution 4.0 International License. ISSAI is a member of DataCite, and a digital object identifier (DOI) was assigned by the ISSAI Repository to the SpeakingFaces dataset 
 ({\url{https://doi.org/10.48333/smgd-yj77}}, accessed April 2, 2021). The database is comprised of 142 subjects in total, with a gender balance of 68 female and 74 male participants, with the ages of participants ranging from 20 to 65, and an average age of 31. The data is split into three parts: train set, validation set, and test set. The subjects and commands in each set are unique, i.e., they are non-overlapping. Table~\ref{table:splitstat} presents the information on the three splits of SpeakingFaces. 
\end{paracol}
   \begin{specialtable}[H]
    \tablesize{\footnotesize}
    \widetable
\caption{\label{table:splitstat} Statistics on SpeakingFaces.}
\setlength{\cellWidtha}{\columnwidth/5-2\tabcolsep+1.6in}
\setlength{\cellWidthb}{\columnwidth/5-2\tabcolsep-.2in}
\setlength{\cellWidthc}{\columnwidth/5-2\tabcolsep-.6in}
\setlength{\cellWidthd}{\columnwidth/5-2\tabcolsep-.6in}
\setlength{\cellWidthe}{\columnwidth/5-2\tabcolsep-.2in}
\scalebox{1}[1]{\begin{tabularx}{\columnwidth}{
>{\PreserveBackslash\raggedright}m{\cellWidtha}
>{\PreserveBackslash\centering}m{\cellWidthb}
>{\PreserveBackslash\centering}m{\cellWidthc}
>{\PreserveBackslash\centering}m{\cellWidthd}
>{\PreserveBackslash\centering}m{\cellWidthe}}
\toprule
\textbf{Category}                                                   & \textbf{Train}    & \textbf{Valid}    & \textbf{Test} & \textbf{Total}\\
\midrule
\# Speakers                                                         & 100           & 20            & 22            & 142\\\midrule         
Speaker IDs                                                         & 1--100         & 101--120       & 121--142       & 1--142\\\midrule   
\# Commands                                                         & 9771         & 1963         & 1977         & 13,711\\\midrule      
\# Unique commands                                                  & 1297         & 250           & 250           & 1797\\\midrule       
\# Words                                                            & 52,769        & 10,324        & 11,177         & 74,270\\\midrule      
\# Unique words                                                     & 683           & 337           & 312	        & 823\\\midrule         
{\# Visual-thermal image pairs for trials with commands}    & 1,054,989     & 223,528       & 234,071       & 1,512,588\\\midrule   
{\# Visual-thermal image pairs for trials w/o commands}     & 1,620,000     &	356,400     & 324,000	    & 2,268,000\\\midrule   
{Duration in hours for trials with commands}                & 16.1          & 3.2           & 3.5           & 22.8\\\midrule        
{Duration in hours for trials w/o commands}                 & 16.1          & 3.2           & 3.5           & 22.8\\\midrule        
Size of raw data in TB                                              & 5.0           & 1.0           & 1.1         & 7.1\\          
\bottomrule
\end{tabularx}}
\end{specialtable}
\begin{paracol}{2}
\switchcolumn

The public repository consists of annotated data (metadata), raw data, and clean data. The repository structure is presented in Figure~\ref{fig:struct}a. Let us first introduce the notation relevant to the names of directories and files in the figure:
\begin{itemize}
    \item subID $= \{1 \dots 142\}$ denotes the subject number. 
    \item trialID $=  \{1,2\}$ denotes the trial number.
    \item sessionID is 1 if the session does not involve utterances and 2 otherwise. 
    \item posID $= \{1 \dots 9\}$ denotes the camera position.
    \item commandID $= \{1 \dots 1,297\}$ denotes the command number.
    \item frameID $= \{1 \dots 900\}$ the number of an image in a sequence.  
    \item streamID is 1 for thermal images, 2 for visual images, and 3 for the aligned version of the visual images.  
    \item micID is 1 for the left microphone and 2 for the right microphone on the web camera.
\end{itemize}

\textls[-12]{The annotated data are stored in the 
 metadata directory, which consists of the subjects.csv file and the commands subdirectory.
The former contains information on the ID, split (train/valid/test), gender, ethnicity, age, and accessories (hat, glasses, etc.) in both trials for each subject. 
The latter consists of sub\_subID\_trial\_trialID.csv,} composed of records on each command uttered by the subject subID in the trial trialID. 
There are 284 files in total, two files for each of the 142 subjects. 
A record includes the command name, the command identifier, the identifier of a camera position (see Figure~\ref{fig:9pos}) at which the utterance was captured, the transcription of the uttered command, and information on the artifacts detected in the recording.  

\end{paracol}
\begin{figure}[H]
\widefigure
\includegraphics[width=0.8\linewidth,trim={0cm -0.1cm 0cm 0.1cm},clip=true]{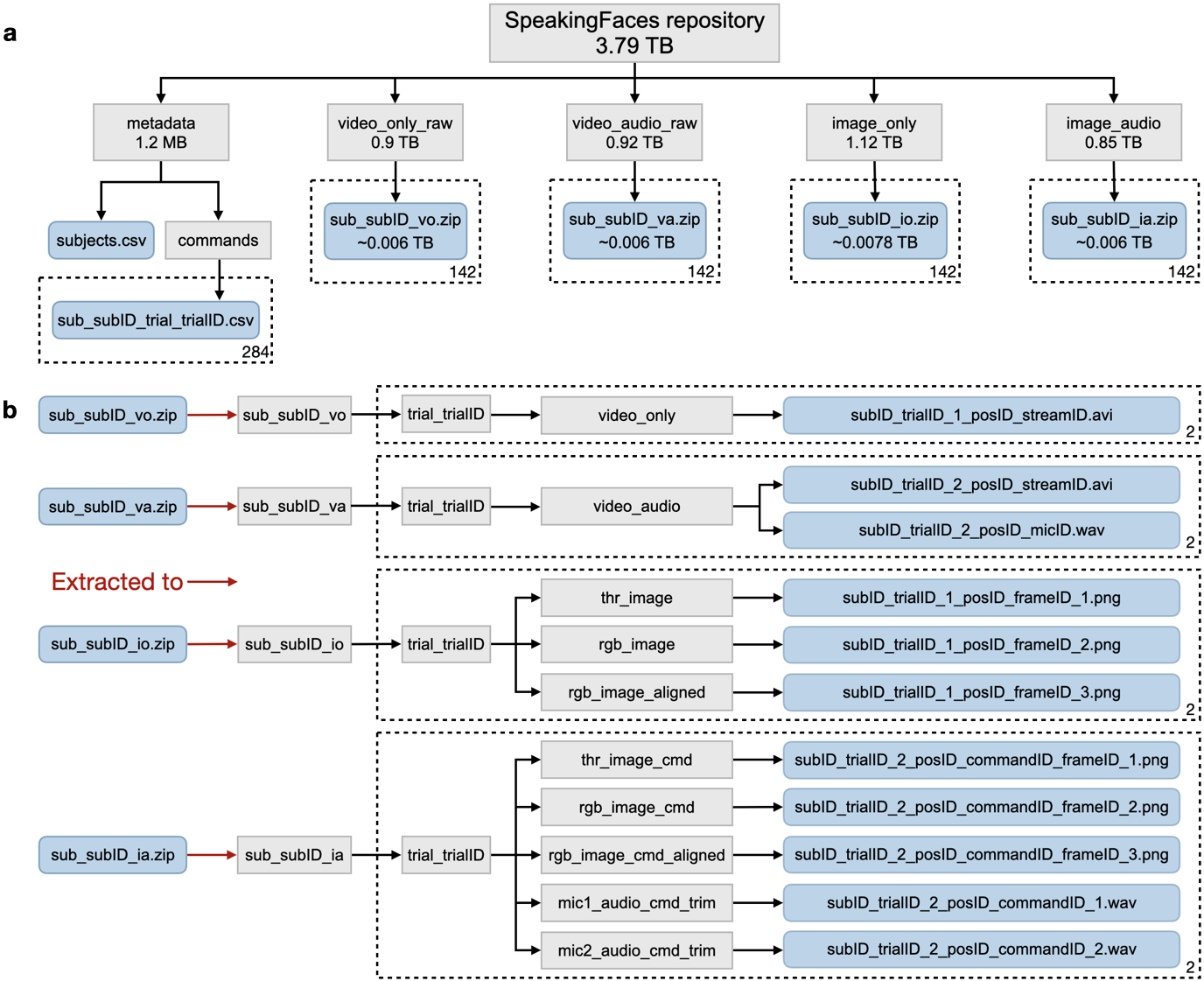}
\caption{(\textbf{a}) File structure of SpeakingFaces repository. File names are suffixed by $subID = \{1, \dots, 142\}$ and\linebreak $trialID = \{1,2\}$, bringing the total number of files to the indicated max (142 or 284). (\textbf{b}) Illustration of the extraction procedure for the archived files in the SpeakingFaces repository and the final output files for each type. Directories are shaded gray, and the files are shaded blue for both subfigures.}
\label{fig:struct}
\end{figure}
\begin{paracol}{2}
\switchcolumn

There are four categories of artifacts, corresponding to the four data streams: thermal, visual, audio, and text. For each stream, Table~\ref{table:art_legend} lists detected artifacts and the corresponding numerical value recorded in the metadata. Thus, an utterance that is “clean” of any noise in the data would have 0 in all four categories. In total, $86\%$ of the utterances are clean of any noise. Depending on the application of the dataset, users can decide which of the artifacts is acceptable and select the data in accordance with their preferences.

  \begin{specialtable}[H]
    \tablesize{\small}
\caption{\label{table:art_legend} Legend for data types and detected artifacts.}
\setlength{\cellWidtha}{\columnwidth/5-2\tabcolsep-.2in}
\setlength{\cellWidthb}{\columnwidth/5-2\tabcolsep-0in}
\setlength{\cellWidthc}{\columnwidth/5-2\tabcolsep-0in}
\setlength{\cellWidthd}{\columnwidth/5-2\tabcolsep+.2in}
\setlength{\cellWidthe}{\columnwidth/5-2\tabcolsep-0in}
\scalebox{1}[1]{\begin{tabularx}{\columnwidth}{
>{\PreserveBackslash\centering}m{\cellWidtha}
>{\PreserveBackslash\centering}m{\cellWidthb}
>{\PreserveBackslash\centering}m{\cellWidthc}
>{\PreserveBackslash\centering}m{\cellWidthd}
>{\PreserveBackslash\centering}m{\cellWidthe}}
\toprule
\textbf{Value} & \textbf{Thermal} & \textbf{Visual} & \textbf{Audio} & \textbf{Text}\\
\midrule
0 & Clean & Clean & Clean & Clean\\ 
1 & Frozen images & Blurred images & Background noise & Minor text noise\\
2 &  - & Cut chin & - & Invalid\\
3 & - & Both 1 and 2 & - & -\\
\bottomrule
\end{tabularx}}
\end{specialtable}

The raw data on the ``non-speaking'' session can be found in video\_only\_raw, which contains the compressed version of unprocessed video files from both trials for a given subject. The raw data for the other session can be located in video\_audio\_raw. Similarly, it consists of compressed and unprocessed video/audio files from both trials for a given subject. The clean data correspond to the result of the whole data preprocessing pipeline (see Figure~\ref{fig:pipeline}). The img\_only directory contains the compressed version of thermal, visual, and aligned visual image frames from the first session. In addition to the image frames, the img\_audio folder contains the audio tracks for each spoken utterance in the second session. The folders video\_only\_raw, video\_audio\_raw, img\_only, img\_audio contain 142 files each. Each file is a .zip archive that contains data for one of the subjects. The data should be extracted first, and the resulting file structure is presented in Figure~\ref{fig:struct}b.
Further details on the database structure and download instructions can be accessed on the repository page 
\linebreak ({\url{https://issai.nu.edu.kz/download-speaking-faces/}}, accessed April 2, 2021).

\section{Results and Discussion}
\label{sec:Preliminary Experiments}
We developed two baseline tasks to demonstrate the utility and reliability of the SpeakingFaces multimodal dataset.
The first task utilizes the three data streams (visual, thermal, and audio) to classify the gender of subjects under clean and noisy environments.
The second task aims to learn a thermal-to-visual image translation model in order to demonstrate a transfer of domain knowledge between the two data streams.

\subsection{Gender Classification} 
The goal of this task is to predict the gender of a subject using the information from a single utterance, consisting of visual, thermal, and audio data streams.
To achieve this goal, we constructed a multimodal gender classification system using our SpeakingFaces dataset.
A successful gender classification system can improve the performance of many applications, including HCI, surveillance and security systems, image/video retrieval, and so on~\cite{rai2012gender}.

The gender classification model is based on LipNet~\cite{DBLP:journals/corr/AssaelSWF16} architecture consisting of two main modules: an encoder and a classifier.
The encoder module is constructed by combining deep convolutional neural networks (CNN) with the stack of bidirectional recurrent neural network (BRNN) layers:

\begin{equation}
Encoder(\cdot) \triangleq BRNN(CNN(\cdot)).
\end{equation}

The encoder module is used to transform an $N$-length input feature sequence\linebreak $X=\{x_1,\dots,x_N\}$ into a hidden feature vector $h$ as follows:
  
\begin{equation}
h = Encoder(X),
\end{equation}
where $x_i$ is an three-dimensional tensor for images or a two-dimensional tensor for the spectrograms generated from the audio records. A separate encoder module is trained for each data stream, producing three hidden vector representations: $h_{visual}$, $h_{thermal}$, and $h_{audio}$. These generated hidden features are then concatenated and fed to the classifier module.

The classifier module consists of two fully-connected layers with the rectified linear unit (\emph{ReLU}) activation and single linear layer followed by the sigmoid activation:

\begin{equation}
Classifier(\cdot) \triangleq Sigmoid(Linear(ReLU(ReLU(\cdot)))),
\end{equation}
where the linear layer is used to convert a vector to a scalar.
The classifier takes the generated hidden features and outputs probability distribution over the two classes\linebreak $y\in \{female,male\}$ as follows:

\begin{align}
h_{visual} &= Encoder_1(X_{visual}) \\
h_{thermal} &= Encoder_2(X_{thermal}) \\
h_{audio} &= Encoder_3(X_{audio}) \\
P(y|X_{visual},X_{thermal},X_{audio}) &= Classifier([h^T_{visual},h^T_{thermal},h^T_{audio}]^T),
\end{align}
where $Encoder_i(\cdot)$ is a $i$-th encoder dedicated to the specific data stream, and $T$ denotes the transpose operation.

The input sequence $X$ is constructed as follows.
For visual and thermal streams, we used the same number of equidistantly spaced frames.
For audio streams, we used mel-spectogram features computed over a 0.4-second snippet extracted from the middle of uttered commands.
To evaluate the robustness of multimodal gender classification model, we constructed noisy versions of input features for the validation and test sets.
The noisy input features $X_{noisy}$ were generated by including additive white Gaussian noise (AWGN):

\begin{equation}
X_{noisy}=X+Z,
\end{equation}
where Z$\sim$$\mathnormal{N}(0,\Sigma)$.
To estimate the noise variance $\Sigma$, we steadily increased it up to the point when the input data were sufficiently corrupted, that is, the gender classifier makes random predictions. 
As a result, the noise variance $\Sigma$ for image and audio streams was set to 100 and 5, respectively.

All models were trained on a single V100 GPU running on the NVIDIA DGX-2 server using the clean training set.
All hyper-parameters were tuned using the clean validation set.
In particular, we optimized model parameters using Adadelta~\cite{zeiler2012adadelta} with the initial learning rate of 0.1 for 200 epochs.
As a regularization, we applied dropout, which was tuned for each model independently.
We set the batch size to 256 and applied gradient clipping with a threshold of 10 to prevent the gradients from exploding.
The best-performing model was evaluated using the clean and noisy versions of the validation and test sets.
The system implementation including the model specifications and other hyper-parameter values are provided in our GitHub repository 
 ({\url{https://github.com/IS2AI/SpeakingFaces/tree/master/baseline_gender}}, accessed February 24, 2021).

The model inference results are given in Table~\ref{table:gender_age_class}.
In these experiments, we set the number of visual and thermal frames to three, extracted from the beginning, middle and end of an utterance. 
We examined different number of frames and observed that three equidistantly spaced frames were sufficient to achieve a good predictive performance, i.e., increasing the number of frames commensurately lengthened both training and inference time, but did not produce any noticeable performance improvement (see Figure~\ref{fig:frames}).
In the best-case scenario, when all of the three data streams are clean (ID 1), the gender classifier achieves the highest accuracy rate of 96\% on the test set.
When all the three data streams are noisy (ID 8), the model performance is random, equivalent to a coin toss.
In other scenarios, when only one or two data streams are corrupted (IDs 2--7), the model achieves an accuracy of 
 65.8--95.6\% on the test set; these results serve to demonstrate the robustness of using multimodal systems.

 \begin{specialtable}[H]
    \tablesize{\small}
\caption{\label{table:gender_age_class} The accuracy of multimodal gender classification model evaluated on the clean and noisy input data streams. `n/a' stands for not available. The best accuracy results for the validation and test sets are boldfaced.}
\setlength{\cellWidtha}{\columnwidth/6-2\tabcolsep+0.0in}
\setlength{\cellWidthb}{\columnwidth/6-2\tabcolsep+0.0in}
\setlength{\cellWidthc}{\columnwidth/6-2\tabcolsep+0.0in}
\setlength{\cellWidthd}{\columnwidth/6-2\tabcolsep+0.0in}
\setlength{\cellWidthe}{\columnwidth/6-2\tabcolsep+0.0in}
\setlength{\cellWidthf}{\columnwidth/6-2\tabcolsep+0.0in}
\scalebox{1}[1]{\begin{tabularx}{\columnwidth}{
>{\PreserveBackslash\centering}m{\cellWidtha}
>{\PreserveBackslash\centering}m{\cellWidthb}
>{\PreserveBackslash\centering}m{\cellWidthc}
>{\PreserveBackslash\centering}m{\cellWidthd}
>{\PreserveBackslash\centering}m{\cellWidthe}
>{\PreserveBackslash\centering}m{\cellWidthf}}
\toprule
\multirow{2}{*}{\vspace{-5pt}\textbf{ID}} & \multicolumn{3}{c}{\textbf{Data streams}}               & \multicolumn{2}{c}{\textbf{Accuracy (\%)}} \\\cmidrule{2-6}
                    & \textbf{Visual}       & \textbf{Audio}            & \textbf{Thermal}  & \textbf{Valid}    & \textbf{Test} \\ 
\midrule
1                   & \multirow{4}{*}{\vspace{-5pt}clean}& \multirow{2}{*}{clean}    & clean         & 89.6                  & \textbf{96.0} 
 \\\cline{4-4}
2                   &                       &                           & noisy         & 87.9                  & 95.6          \\\cmidrule{3-4}
3                   &                       & \multirow{2}{*}{\vspace{-5pt}noisy}    & clean         & 81.9                  & 84.4          \\\cmidrule{4-4}
4                   &                       &                           & noisy         & 76.6                  & 82.0          \\\cmidrule{2-4}
5                   & \multirow{4}{*}{\vspace{-12pt}noisy}& \multirow{2}{*}{\vspace{-5pt}clean}    & clean         & 89.2                  & 93.6          \\\cmidrule{4-4}
6                   &                       &                           & noisy         & 84.5                  & 88.2          \\\cmidrule{3-4}
7                   &                       & \multirow{2}{*}{\vspace{-5pt}noisy}    & clean         & 67.5                  & 65.8          \\\cmidrule{4-4}
8                   &                       &                           & noisy         & 55.2                  & 50.1          \\\cmidrule{2-4}
9                   & clean                 & n/a                       & n/a           & 91.5                  & 91.3          \\\cmidrule{2-4}
10                  & n/a                   & clean                     & n/a           & 90.1                  & 94.8          \\\cmidrule{2-4}
11                  & n/a                   & n/a                       & clean         & \textbf{94.8}         & 94.3          \\
\bottomrule
\end{tabularx}}
\end{specialtable}
\vspace{-6pt}
\begin{figure}[H]
\includegraphics[width=0.7\linewidth]{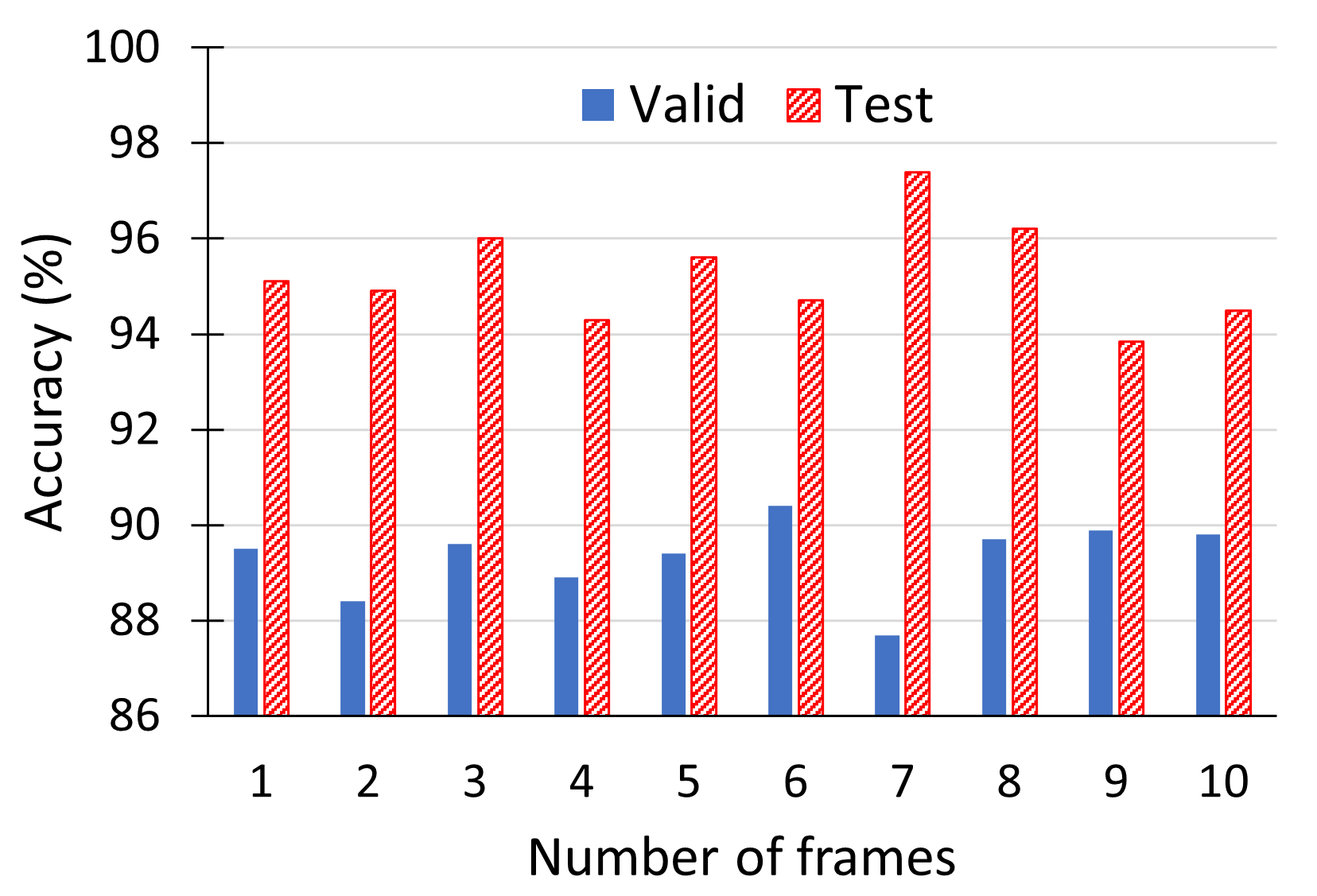}
\caption{The impact of increasing the number of visual and thermal frames on the multimodal gender classification accuracy, when all streams are clean.}
\label{fig:frames}
\end{figure}

The experiment results show that the most informative data stream is the audio, followed by the visual and then thermal stream.
When considering the case where only a single stream is noisy, the corruption of the audio stream drops the accuracy rate by 11.6\% (ID~1 vs. ID~3), whereas for the visual and thermal streams, the accuracy drops by 2.4\% (ID~1 vs. ID~5) and 0.4\% (ID~1 vs. ID~2), respectively.
Now, considering the case where two streams are noisy: when the audio (ID~6) stream is clean (and the others corrupted), the accuracy is 88.2\%, while, when only the visual (ID~4) and thermal (ID~7) images are clean, the performances are 82.0\% and 65.7\%, respectively.
We presume that during the training phase, the multimodal model decides to emphasize the audio features such that the relative contributions of the visual and thermal streams are de-emphasized. Presumably, this issue can be addressed by using attention-based models~\cite{DBLP:journals/corr/BahdanauCB14}.
Although the thermal stream seems to be relatively less consequential, it is still extremely useful in the case where the visual stream is corrupted (e.g., at night), where 5.4\% of improvement on the test set is gained (ID 5 vs. ID 6).
The experimental results successfully demonstrate the advantages of examining multiple data streams, and the utility of the SpeakingFaces dataset. We believe that the gender classification model can achieve even better results, with further development of the architectural structure and tuning of the hyper-parameter values, though this optimization work lies beyond the scope of this baseline example.

To further verify the reliability of the SpeakingFaces dataset, we evaluated the performance of each data stream independently.
Specifically, we trained a gender classification model using only a single data stream.
The model architecture was same as in the previous experiment setup, except that the number of encoders was reduced from three to one.
This experiment was conducted using only the clean version of the data.
The obtained results (IDs~9--11) show that all the data streams achieve an accuracy score of above 90\% on both validation and test sets.
The best accuracy on the test set is achieved by the model trained on the audio (ID~10) stream, followed by the thermal (ID~11) and visual (ID~9) streams.
These experimental results demonstrate the reliability of each data stream present in the SpeakingFaces dataset.

As was previously mentioned, the gender classification experiments were conducted to demonstrate the utility and trustworthiness of the available modalities in the SpeakingFaces. In particular, the multimodal experiments were conducted to demonstrate the robustness of the recognition system trained on the three streams under different conditions. On the other hand, the unimodal experiments were conducted to show the reliability of each individual stream present in the dataset. These experiments are not intended to compare unimodal versus multimodal systems, they were generated as a proof-of-concept. Further investigation on hyper-parameter tuning and architectural search to improve and compare the performance of unimodal and multimodal models is underway as a separate contribution.

\subsection{Thermal-to-Visual Facial Image Translation}
Facial features which are distinctly discernible in the visible images are not clearly observable in the corresponding thermal versions (see Figure~\ref{fig:9pos}). As a result, models developed for visual images (e.g., facial landmark detection, face recognition) cannot be utilized directly on thermal images. Therefore, in this task, we aim to address the problem of generating a realistic visual-spectrum version of a given thermal facial image.

Generative Adversarial Networks (GANs)~\cite{NIPS2014_5423} have been successfully deployed for generating realistic images; in particular, Pix2Pix~\cite{isola2017image}, CycleGAN~\cite{zhu2017unpaired}, and CUT~\cite{park2020contrastive} have been shown to produce promising results in translating images from one domain to another. 
Zhang et al. introduced a Pix2Pix-based approach that focused on achieving a high face recognition accuracy of their generated visible images by incorporating an explicit closed-set face recognition loss~\cite{zhang2018tv}. However, their image output lacked distinct facial features and high image quality, which was the priority of Wang et al.~\cite{wang2018thermal}. They combined CycleGAN with a new detector network that located facial landmarks in generated visible images and aimed to guide the generator in producing realistic results. Both works were impaired by the relatively small number of image pairs and the use of low resolution thermal cameras. Zhang et al. filtered the IRIS dataset~\cite{iris} down to 695 image pairs, and Wang et al. collected 792 image pairs using FLIR AX5 thermal camera with a resolution of $320\times256$. The latter dataset is not publicly available. 

In our case, we experimented with CycleGAN and CUT to map thermal faces to visual-spectrum. The SpeakingFaces contains images of 142 subjects; 100 subjects were used for training and 42 were left for testing. We used the second session data, where participants uttered commands, and randomly selected three images for every position of each subject, which resulted in 2700 and 1134 thermal-visual image pairs for training and testing, respectively. To prepare the experimental data, we utilized the OpenCV’s deep learning face detector~\cite{github_face_detector} to identify faces in visible images. Noting that the thermal and visual images are aligned, we used the bounding boxes extracted from the visible images to delineate faces in both image streams. In cases where faces were not detected, we manually specified the coordinates of the bounding boxes. The instructions on how to access this version of SpeakingFaces can be found in our Github repository (\url{https://github.com/IS2AI/SpeakingFaces/tree/master/baseline_domain_transfer}, accessed March 11, 2021).

All models were trained on a single V100 GPU running on the NVIDIA DGX-2 server using the training set. For both CycleGAN and CUT, the generator architecture was comprised of ResNet-9 blocks, trained using identical hyperparameter values with a batch size of 1, an image load size of 130, and an image crop size of 128. The rest of the training and testing details can be accessed in our GitHub repository (\url{https://github.com/IS2AI/SpeakingFaces/tree/master/baseline_domain_transfer}, accessed March 11, 2021). 

We used two methods to quantitatively assess our experimental results. The first one was the Fréchet inception distance (FID) metric that compares the distribution of generated images with the distribution of real images~\cite{heusel2017gans}. The second method is based on the dlib's face recognition model~\cite{dlib09, github_face_recognition}, which was trained on visual images, to show accuracy metrics on real visual, generated visual, and real thermal images from the test set. 

The recognition model extracts a 128-dimension encoding for a given facial image and matches faces by comparing the Euclidean distance between the encodings. We started with the real visual images from the first trial to get the ground truth features. To do so, we built a feature matrix $X\in R^{1134\times128}$ by extracting face encodings from the first trial data, where the columns represent features and the rows represent image samples. We also saved the corresponding labels (a numeric identifier of each subject) in the vector $y \in R^{1134}$. 

Next, we used the second trial images (real visual, real thermal, generated visual CycleGAN, and generated visual CUT) to evaluate the model performance. We computed encodings for each image in the second trial and calculated the Euclidean distance with every feature vector from $X$. If the distance was below a predefined threshold, then we had a match. Note, $X$ contains 27 (three images from each of the nine positions) embedding vectors for each subject, so when we compared each face in the second trial with the encodings in $X$, we chose the label with the highest number of matches. The implementation of the face recognition pipeline can be found in our GitHub repository\linebreak (\url{https://github.com/IS2AI/SpeakingFaces/tree/master/baseline_domain_transfer}, accessed March 11, 2021). 

The threshold value, or the tolerance, was tuned to meet the precision/recall trade-off on real visual images.
The larger value increases a number of false positive predictions, while the lower value leads to a higher count of false negative predictions. The threshold value for our data was established at 0.45, to better balance the precision/recall trade-off. 

A subset of generated images is presented in Figure~\ref{fig:cyclegan}; the rest can be found in our Github repository (\url{https://github.com/IS2AI/SpeakingFaces/tree/master/baseline_domain_transfer } Accessed March 11, 2021). Accessed March 11, 2021). Compared to the images generated by CUT, the output of CycleGAN is of much higher quality. The CycleGAN images are close to the target visible images not only in the structure of facial features, but also in the overall appearance for a variety of head postures. The model produced samples with smoother and more coherent skin texture and color. Overall, the hair is realistically drawn, though both models were biased towards brown-haired individuals, so they failed to provide the right hair color for subject ID 1. Interestingly, both learned to correctly predict the gender of each person; for example, the generators drew facial hair for the male subjects.

The qualitative assessment of the synthesized images is supported by the FID metric and face recognition results for both models. The FID scores were 22.12 for CUT and 18.95 for CycleGAN. This means that the CycleGAN-generated images were more similar to real visual images than the ones generated by CUT. The reason might be that, in the training procedure of the CUT model, each patch in the output image should reflect the content of the corresponding patch in the input image, whereas the CycleGAN enforces a cycle consistency between entire images. 
\begin{figure}[H]
    \includegraphics[width=.95\linewidth,
    trim={0cm -1.0cm 0cm -0.75cm},
    clip=true]{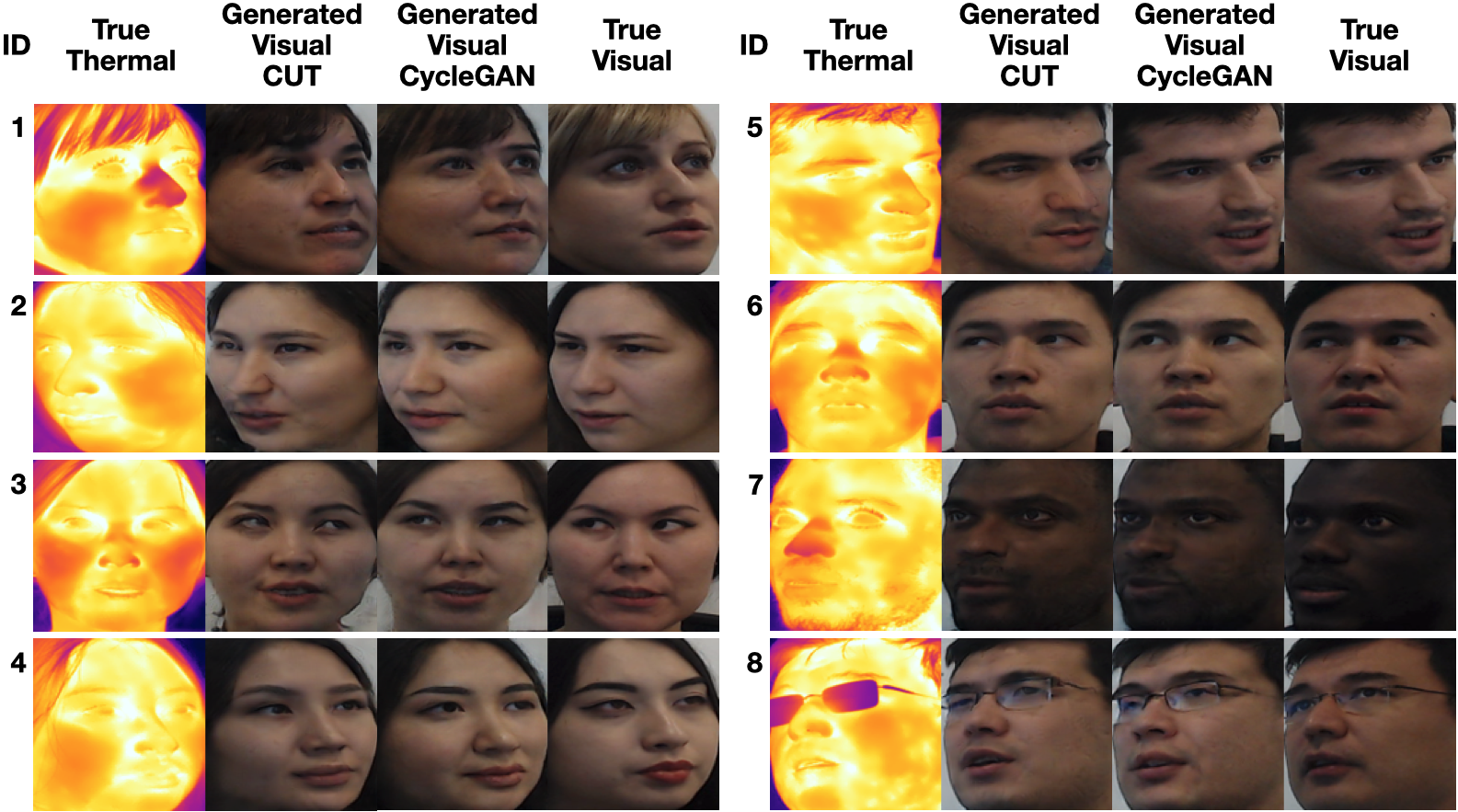}
\caption{Thermal-to-visual image translation results using CycleGAN. For each column, left to right: real thermal image; generated visual image by CUT; generated visual image by CycleGAN; and real visual image.}
\label{fig:cyclegan}
\end{figure}

The face recognition results are shown in Table~\ref{table:face_rec}. As expected, the best outcomes were obtained from the real visual images, while the worst were from the real thermal images, because the deployed recognition model was trained on visual images. The results of the CycleGAN model are noticeably better than those of the CUT model; this is also supported by their FID scores and our qualitative examination.
The quality of the generated images requires further improvement as compared to the outcomes achieved with the real visible images.
We hypothesize that the realism of the output of these models was affected by the following factors:
\begin{itemize}
    \item The model may be biased towards young people, due to the observation that $34\%$ of participating subjects were 20--25 years old. As a result, the model in some cases generated a younger version of the subject.
    \item The model may be biased towards Asian people, given that the majority of the participating subjects were Asians. As an example, in the case of some subjects wearing glasses, the depiction of eyes seems skewed towards an Asian presentation.
\end{itemize}

Even taking into account the noted slight biases, the recognition accuracy on the generated images is significantly higher than that on the real thermal images. These results showcase that SpeakingFaces can indeed be utilized for image translation tasks, and we encourage other researchers to experiment further and compare their results.  

\end{paracol}
 \begin{specialtable}[H]
    \tablesize{\small}
    \widetable
\caption{\label{table:face_rec} The results of the face recognition model. TP---the number of true positives. FP---the number of false positives. FN---the number of false negatives.}
\setlength{\cellWidtha}{\columnwidth/8-2\tabcolsep+1in}
\setlength{\cellWidthb}{\columnwidth/8-2\tabcolsep-.2in}
\setlength{\cellWidthc}{\columnwidth/8-2\tabcolsep+0.0in}
\setlength{\cellWidthd}{\columnwidth/8-2\tabcolsep-.4in}
\setlength{\cellWidthe}{\columnwidth/8-2\tabcolsep-.2in}
\setlength{\cellWidthf}{\columnwidth/8-2\tabcolsep-.2in}
\setlength{\cellWidthg}{\columnwidth/8-2\tabcolsep+0.0in}
\setlength{\cellWidthh}{\columnwidth/8-2\tabcolsep+0.0in}
\scalebox{1}[1]{\begin{tabularx}{\columnwidth}{
>{\PreserveBackslash\centering}m{\cellWidtha}
>{\PreserveBackslash\centering}m{\cellWidthb}
>{\PreserveBackslash\centering}m{\cellWidthc}
>{\PreserveBackslash\centering}m{\cellWidthd}
>{\PreserveBackslash\centering}m{\cellWidthe}
>{\PreserveBackslash\centering}m{\cellWidthf}
>{\PreserveBackslash\centering}m{\cellWidthg}
>{\PreserveBackslash\centering}m{\cellWidthh}}
\toprule
\textbf{Domain}             & \textbf{Tolerance}& \textbf{TP}   & \textbf{FP}   & \textbf{FN}   & \textbf{Total}& \textbf{Precision}& \textbf{Recall} \\\midrule
Visual (real)               & 0.45              & 1110          & 4             & 20            & 1134          & 0.99              & 0.98 \\
Thermal (real)              & 0.45              & 12            & 555           & 567           & 1134          & 0.02              & 0.02 \\
CUT (generated visual)      & 0.45              & 236           & 382           & 516           & 1134          & 0.38              & 0.31 \\
CycleGAN (generated visual) & 0.45              & 397           & 317           & 420           & 1134          & 0.56              & 0.49 \\
\bottomrule
\end{tabularx}}
\end{specialtable}
\begin{paracol}{2}
\switchcolumn

\subsection{Limitations}
The SpeakingFaces dataset was acquired in a semi-controlled laboratory setting, which may present certain limitations to the work when used in unconstrained real-world settings where there is less control over camera angles, distance, lighting, and temperature. The first limitation entails the orientation of the subject to the camera. We used nine camera positions, though in an open setting it is likely that a wider range of facial poses would be encountered. The second limitation involves the distance of the subject from the camera: the distance did not vary in the laboratory setting. In an open setting, the distance could vary considerably, which could result in reduced resolution of facial images, thus diminishing the accuracy of the results. The third limitation is that our dataset was acquired under consistent illumination and temperature conditions. In a real-world deployment there could be wide variation in the surrounding thermal conditions, ambient light intensity and illumination directions. To address these issues, as future work, it is proposed to enhance the dataset with the acquisition of in-the-wild subject data. The models trained on the original dataset could be further fine-tuned with the real-world dataset using transfer~learning.

Another limitation arises from the proposed method of aligning visual images to their thermal pairs. Our method (as described in Section \ref{sec:Data Preprocessing}) was based on planar homography and ArUco markers. Since the corners of the marker might not be detected very accurately in the thermal image due to heat dissipation, we estimated the averaged value of the homography matrix by collecting ArUco marker images from different positions and orientations. The averaged homography matrix allowed us to align well in terms of scale and position, but not in terms of orientation. 

Despite the large size of the dataset, it might be insufficient to build robust multimodal models for the tasks, such as speech recognition and lip reading. These tasks require a substantial amount of annotated data, which is expensive and time-consuming to acquire. However, our dataset can be used to fine-tune unimodal models pre-trained on large single stream datasets, as was done in~\cite{shon2020multimodal}.

Lastly, as noted above, the manual operation of the camera introduced variability in the acquisition of visual and thermal data. Nevertheless, we think that such an approach is suitable for the potential deployment of applications built with SpeakingFaces. As previously mentioned, smartphones will likely be the first devices to deploy applications utilizing all the three data streams. These devices are commonly handheld, thus it will be more suitable to train models on the data that were collected in a similar manner. Furthermore, manual operations introduce variability in framing and thereby improve the robustness of subsequent machine learning applications.

\section{Conclusions}
\label{sec:Conclusion}
We introduce SpeakingFaces as a large-scale multimodal dataset to extend existing research in the general areas of HCI, biometric authentication, and recognition systems. SpeakingFaces consists of synchronized audio, thermal, and visual streams gathered from a diverse population of subjects. 

To demonstrate the utility, we applied our data to thermal-to-visible image translation and multimodal gender classification using thermal, visible, and audio data streams. 
Based on the experimental results, we see that SpeakingFaces has the following positive impacts. 
First, it enables in-depth research in the areas of multimodal recognition systems using visual, thermal, and audio modalities.
Second, the large number of samples in the dataset enables the construction and study of data-hungry algorithms involving neural networks.
Lastly, synchronized multimodal data can open up new opportunities for research in domain transfer.

In future work, we plan to utilize our dataset in other multimodal tasks, such as audio--visual--thermal speech and speaker recognition. We also plan to annotate the thermal data with facial landmarks to build a landmark detection model that can be deployed for face alignment in face recognition, vital sign recognition, and drowsiness detection. We also intend to create an additional in-the-wild version of SpeakingFaces, to overcome the noted limitations of the original dataset attributed to the semi-controlled laboratory collection setting. Considering that smartphones and other intelligent devices can be potentially integrated with additional sensors, such as high-speed, depth, and event-based cameras, the SpeakingFaces dataset can be expanded with these modalities.


\vspace{6pt} 



\authorcontributions{Conceptualization, M.A. and H.A.V.; methodology, M.A., A.K., and Y.K.; software, M.A., S.J., and A.K.; validation, A.K. and Y.K.; formal analysis, M.A., A.K., and Y.K.; investigation, M.A. and S.J.; resources, M.L. and H.A.V.; data curation, M.A., A.K., and S.J.; writing---original draft preparation, M.A.; writing---review and editing, M.L. and H.A.V.; visualization, M.A. and A.K.; supervision, H.A.V.; project administration, H.A.V.; funding acquisition, H.A.V. All authors have read and agreed to the published version of the manuscript.}

\funding{This research received no external funding.}

\institutionalreview{The study was conducted according to the guidelines of the Declaration of Helsinki, and approved by the Institutional Research Ethics Committee of Nazarbayev University (14 October 2019).}

\informedconsent{Informed consent was obtained from all subjects involved in the~study.}

\dataavailability{The data presented in this study are openly available on our local storage servers at 
 \url{https://doi.org/10.48333/smgd-yj77}, accessed April 2, 2021.} 

\acknowledgments{We would like to express our gratitude to Aknur Karabay, Azamat Sarkytbayev, Saida Mussakhojayeva, Raushan Utemuratova, Mukhamet Nurpeiissov, Zhaniya Koishybaeva, and Nazym Shakirkhozha for assisting us in processing the dataset.
We also thank our technical editor Rustem Yeshpanov, video editor Dias Aitkenov, computer engineer Almas Mirzakhmetov, and project coordinator Yerbol Absalyamov for their help with other administrative and technical tasks.}

\conflictsofinterest{The authors declare no conflict of interest.}



\abbreviations{Abbreviations}{
The following abbreviations are used in this manuscript:\\

\noindent 
\begin{tabular}{@{}ll}
AWGN & Additive White Gaussian Noise\\
BRNN & Bidirectional Recurrent Neural Networks\\
CNN & Convolutional Neural Networks\\
DOI & Digital Object Identifier\\
FID & Fréchet Inception Distance \\
FPS & Frames per Second \\
GAN & Generative Adversarial Networks\\
HCI & Human-Computer Interaction\\
ID & Identifier \\
IoT & Internet of Things \\
ISSAI & Institute of Smart Systems and Artificial Intelligence\\
LRS & Lip Reading Sentences\\
LRW & Lip Reading in the Wild\\
ReLU & Rectified Linear Unit\\
RoI & Region-of-Interest
\end{tabular}}

\end{paracol}
\reftitle{References}






\begin{thebibliography}{999}

\bibitem[Chen \em{et~al.}(2020)Chen, Wang, and Qian]{chen2020multi}
Chen, Z.; Wang, S.; Qian, Y.
\newblock Multi-modality Matters: A Performance Leap on VoxCeleb.
\newblock  In Proceedings of the Interspeech, Online, 25--29 October 2020; pp. 2252--2256.

\bibitem[Gade and Moeslund(2014)]{gade2014thermal}
Gade, R.; Moeslund, T.B.
\newblock Thermal cameras and applications: A survey.
\newblock {\em Mach. Vis. Appl.} {\bf 2014}, {\em 25},~245--262.

\bibitem[{Shon} \em{et~al.}(2019){Shon}, {Oh}, and {Glass}]{8683477}
{Shon}, S.; {Oh}, T.; {Glass}, J.
\newblock Noise-tolerant Audio-visual Online Person Verification Using an
  Attention-based Neural Network Fusion.
\newblock  In Proceedings of the ICASSP 2019---2019 IEEE International Conference on Acoustics,
  Speech and Signal Processing (ICASSP), Brighton, UK, 12--17 May 2019; pp. 3995--3999.
\newblock
  doi:{\changeurlcolor{black}\href{https://doi.org/10.1109/ICASSP.2019.8683477}{\detokenize{10.1109/ICASSP.2019.8683477}}}.

\bibitem[Afouras \em{et~al.}(2018)Afouras, Chung, Senior, Vinyals, and
  Zisserman]{afouras2018deep}
Afouras, T.; Chung, J.S.; Senior, A.; Vinyals, O.; Zisserman, A.
\newblock Deep audio-visual speech recognition.
\newblock {\em IEEE Trans. Pattern Anal. Mach. Intell.}
  {\bf 2018}. doi:10.1109/TPAMI.2018.2889052.

\bibitem[Tao and Busso(2021)]{DBLP:journals/tmm/TaoB21}
Tao, F.; Busso, C.
\newblock End-to-End Audiovisual Speech Recognition System With Multitask
  Learning.
\newblock {\em {IEEE} Trans. Multim.} {\bf 2021}, {\em 23},~1--11.
\newblock
  doi:{\changeurlcolor{black}\href{https://doi.org/10.1109/TMM.2020.2975922}{\detokenize{10.1109/TMM.2020.2975922}}}.

\bibitem[{FLIR ONE Pro}(Accessed: February 24, 2021)]{one_pro}
{FLIR ONE Pro}.
\newblock Available online: \url{https://www.flir.com/products/flir-one-pro/} (accessed on 24 February 2021).

\bibitem[{CAT S62 Pro}(Accessed: February 24, 2021)]{s62}
{CAT S62 Pro}.
\newblock Available online: \url{https://www.catphones.com/en-dk/cat-s62-pro-smartphone/} (accessed on 24 February 2021).

\bibitem[{Lepton - LWIR Micro Thermal Camera Module}(Accessed: February 24,
  2021)]{lepton}
{Lepton---LWIR Micro Thermal Camera Module}.
\newblock Available online: \url{https://www.flir.com/products/lepton/} (accessed on 24 February 2021).

\bibitem[Espinosa-Duró \em{et~al.}(2013)Espinosa-Duró, Faundez-Zanuy, and
  Mekyska]{carl}
Espinosa-Duró, V.; Faundez-Zanuy, M.; Mekyska, J.
\newblock A New Face Database Simultaneously Acquired in Visible, Near-Infrared
  and Thermal Spectrums.
\newblock {\em Cogn. Comput.} {\bf 2013}, {\em 5},~119--135.
\newblock
  doi:{\changeurlcolor{black}\href{https://doi.org/10.1007/s12559-012-9163-2}{\detokenize{10.1007/s12559-012-9163-2}}}.

\bibitem[{M}allat and {D}ugelay(2018)]{Mallat18}
{M}allat, K.; {D}ugelay, J.L.
\newblock A benchmark database of visible and thermal paired face images across
  multiple variations.
\newblock  In Proceedings of the International Conference of the Biometrics Special Interest Group,
  {BIOSIG} 2018, Darmstadt, Germany, 26--28 September 2018;
  pp. 199 -- 206.

\bibitem[Hammoud(Accessed: Jan 20, 2020)]{iris}
Hammoud, R.I.
\newblock {IEEE OTCBVS WS Series Bench}.
\newblock Available online: \url{http://vcipl-okstate.org/pbvs/bench/} (accessed on 20 January 2020).

\bibitem[{Wang} \em{et~al.}(2010){Wang}, {Liu}, {Lv}, {Lv}, {Wu}, {Peng},
  {Chen}, and {Wang}]{5523955}
{Wang}, S.; {Liu}, Z.; {Lv}, S.; {Lv}, Y.; {Wu}, G.; {Peng}, P.; {Chen}, F.;
  {Wang}, X.
\newblock A Natural Visible and Infrared Facial Expression Database for
  Expression Recognition and Emotion Inference.
\newblock {\em IEEE Trans. Multimed.} {\bf 2010}, {\em 12},~682--691.
\newblock
  doi:{\changeurlcolor{black}\href{https://doi.org/10.1109/TMM.2010.2060716}{\detokenize{10.1109/TMM.2010.2060716}}}.

\bibitem[Panetta \em{et~al.}(2018)Panetta, Wan, Agaian, Rajeev, Kamath,
  Rajendran, Rao, Kaszowska, Taylor, Samani, et~al.]{panetta2018comprehensive}
Panetta, K.; Wan, Q.; Agaian, S.; Rajeev, S.; Kamath, S.; Rajendran, R.; Rao,
  S.P.; Kaszowska, A.; Taylor, H.A.; Samani, A.; et al.
\newblock A comprehensive database for benchmarking imaging systems.
\newblock {\em IEEE Trans. Pattern Anal. Mach. Intell.}
  {\bf 2018}, {\em 42},~509--520.

\bibitem[Ghiass \em{et~al.}(2018)Ghiass, Bendada, and
  Maldague]{ghiass2018universite}
Ghiass, R.S.; Bendada, H.; Maldague, X.
\newblock \emph{Universit{\'e} Laval Face Motion and Time-Lapse Video Database
  (UL-FMTV)};
\newblock Technical Report; Universit{\'e} Laval: Québec, QC, Canada, 2018.

\bibitem[Poster \em{et~al.}(2021)Poster, Thielke, Nguyen, Rajaraman, Di,
  Fondje, Patel, Short, Riggan, Nasrabadi, et~al.]{poster2021large}
\textls[-12]{Poster, D.; Thielke, M.; Nguyen, R.; Rajaraman, S.; Di, X.; Fondje, C.N.;
  Patel, V.M.; Short, N.J.; Riggan, B.S.; Nasrabadi, N.M.; et al.}
\newblock A Large-Scale, Time-Synchronized Visible and Thermal Face Dataset.
\newblock In Proceedings of the IEEE/CVF Winter Conference on Applications of
  Computer Vision,  Online, 2021; pp. 1559--1568.

\bibitem[Cooke \em{et~al.}(2006)Cooke, Barker, Cunningham, and
  Shao]{cooke2006audio}
Cooke, M.; Barker, J.; Cunningham, S.; Shao, X.
\newblock An audio-visual corpus for speech perception and automatic speech
  recognition.
\newblock {\em  J. Acoust. Soc. Am.} {\bf 2006},
  {\em 120},~2421--2424.

\bibitem[Chung and Zisserman(2016)]{Chung16}
Chung, J.S.; Zisserman, A.
\newblock Lip Reading in the Wild.
\newblock  In Proceedings of the Asian Conference on Computer Vision, Taipei, Taiwan, 20--24 November 2016.

\bibitem[Chung \em{et~al.}(2017)Chung, Senior, Vinyals, and
  Zisserman]{chung2017lip}
Chung, J.S.; Senior, A.; Vinyals, O.; Zisserman, A.
\newblock Lip reading sentences in the wild.
\newblock   In Proceedings of the IEEE Conference on Computer Vision and Pattern
  Recognition (CVPR), Honolulu, HI, USA,  21--26 July 2017; pp. 3444--3453.

\bibitem[Campagna \em{et~al.}(2017)Campagna, Ramesh, Xu, Fischer, and
  Lam]{campagna2017almond}
Campagna, G.; Ramesh, R.; Xu, S.; Fischer, M.; Lam, M.S.
\newblock Almond: The architecture of an open, crowdsourced,
  privacy-preserving, programmable virtual assistant.
\newblock   In Proceedings of the International Conference on World Wide Web, Perth, Australia, 3--7 April 2017; pp.
  341--350.

\bibitem[{The best Siri commands for iOS and MacOS}(Accessed: Dec 20,
  2019)]{siri1}
{The Best Siri Commands for iOS and MacOS}.
\newblock Available online: \url{https://www.digitaltrends.com/mobile/best-siri-commands/} (accessed on 20 December 2019).

\bibitem[{The complete list of Siri commands}(Accessed: Dec 22, 2019)]{siri2}
{The Complete List of Siri Commands}.
\newblock
  Available online: \url{https://www.cnet.com/how-to/the-complete-list-of-siri-commands/} (accessed on 20 December 2019).

\bibitem[Garrido-Jurado \em{et~al.}(2014)Garrido-Jurado, Mu{\~n}oz-Salinas,
  Madrid-Cuevas, and Mar{\'\i}n-Jim{\'e}nez]{garrido2014automatic}
Garrido-Jurado, S.; Mu{\~n}oz-Salinas, R.; Madrid-Cuevas, F.J.;
  Mar{\'\i}n-Jim{\'e}nez, M.J.
\newblock Automatic generation and detection of highly reliable fiducial
  markers under occlusion.
\newblock {\em Pattern Recognit.} {\bf 2014}, {\em 47},~2280--2292.

\bibitem[{ROS Toolbox for MATLAB}(Accessed: January 12, 2020)]{ros}
{ROS Toolbox for MATLAB}.
\newblock Available online: \url{https://www.mathworks.com/products/ros.html} (accessed on 12 January 2020).

\bibitem[{Structural Similarity Index}(Accessed: April 15, 2020)]{ssim_skimage}
{Structural Similarity Index}.
\newblock
  Available online: \url{https://scikit-image.org/docs/dev/auto_examples/transform/plot_ssim.html/} (accessed on 15 April 2020).

\bibitem[{How-To: Python Compare Two Images}(Accessed: April 15,
  2020)]{compare_img}
{How-To: Python Compare Two Images}.
\newblock
  Available online: \url{https://www.pyimagesearch.com/2014/09/15/python-compare-two-images/} (accessed on 15 April 2020).

\bibitem[{Blur detection with OpenCV}(Accessed: April 8, 2020)]{blur_opencv}
{Blur Detection with OpenCV}.
\newblock
  Available online: \url{https://www.pyimagesearch.com/2015/09/07/blur-detection-with-opencv/} (accessed on 8 April 2020).

\bibitem[King(2009)]{dlib09}
King, D.E.
\newblock Dlib-ml: A Machine Learning Toolkit.
\newblock {\em J. Mach. Learn. Res.} {\bf 2009}, {\em
  10},~1755--1758.

\bibitem[Szeliski(2010)]{szeliski2010computer}
Szeliski, R.
\newblock {\em Computer Vision: Algorithms and Applications}; Springer Science
  \& Business Media:  Berlin/Heidelberg, Germany, 
 2010.

\bibitem[{Camera calibration With OpenCV}(Accessed: Dec 5, 2019)]{chess_opencv}
{Camera calibration With OpenCV}.
\newblock
  Available online: \url{https://docs.opencv.org/2.4/doc/tutorials/calib3d/camera_calibration/camera_calibration.html} (accessed on 5 December 2019).

\bibitem[Hwang \em{et~al.}(2015)Hwang, Park, Kim, Choi, and
  So~Kweon]{hwang2015multispectral}
Hwang, S.; Park, J.; Kim, N.; Choi, Y.; So~Kweon, I.
\newblock Multispectral pedestrian detection: Benchmark dataset and baseline.
\newblock   In Proceedings of the IEEE Conference on Computer Vision and Pattern
  Recognition (CVPR), Boston, MA, USA, 7--12 June 2015; pp.~1037--1045.

\bibitem[{Kuzdeuov} \em{et~al.}(2020){Kuzdeuov}, {Rubagotti}, and
  {Varol}]{kuzdeuov2020neural}
{Kuzdeuov}, A.; {Rubagotti}, M.; {Varol}, H.A.
\newblock Neural Network Augmented Sensor Fusion for Pose Estimation of
  Tensegrity Manipulators.
\newblock {\em IEEE Sens. J.} {\bf 2020}.
\newblock
  doi:{\changeurlcolor{black}\href{https://doi.org/10.1109/JSEN.2019.2959574}{\detokenize{10.1109/JSEN.2019.2959574}}}.

\bibitem[Babinec \em{et~al.}(2014)Babinec, Jurišica, Hubinský, and
  Duchoň]{BABINEC20141}
Babinec, A.; Jurišica, L.; Hubinský, P.; Duchoň, F.
\newblock Visual Localization of Mobile Robot Using Artificial Markers.
\newblock {\em Procedia Eng.} {\bf 2014}, {\em 96},~1--9.
\newblock 
  doi:{\changeurlcolor{black}\href{https://doi.org/https://doi.org/10.1016/j.proeng.2014.12.091}{\detokenize{10.1016/j.proeng.2014.12.091}}}.

\bibitem[Bacik \em{et~al.}(2017)Bacik, Durovsky, Fedor, and
  Perdukova]{bacik2017}
Bacik, J.; Durovsky, F.; Fedor, P.; Perdukova, D.
\newblock Autonomous flying with quadrocopter using fuzzy control and ArUco
  markers.
\newblock {\em Intell. Serv. Robot.} {\bf 2017}, {\em 10},~185--194.
\newblock
  doi:{\changeurlcolor{black}\href{https://doi.org/10.1007/s11370-017-0219-8}{\detokenize{10.1007/s11370-017-0219-8}}}.

\bibitem[{Lupu} \em{et~al.}(2017){Lupu}, {Herghelegiu}, {Botezatu},
  {Moldoveanu}, {Ferche}, {Ilie}, and {Levinta}]{lupu2017}
{Lupu}, R.G.; {Herghelegiu}, P.; {Botezatu}, N.; {Moldoveanu}, A.; {Ferche},
  O.; {Ilie}, C.; {Levinta}, A.
\newblock Virtual reality system for stroke recovery for upper limbs using
  ArUco markers.
\newblock   In Proceedings of the International Conference on System Theory, Control and
  Computing (ICSTCC), Sinaia, Romania, 19--21 October 2017; pp. 548--552.
\newblock
  doi:{\changeurlcolor{black}\href{https://doi.org/10.1109/ICSTCC.2017.8107092}{\detokenize{10.1109/ICSTCC.2017.8107092}}}.

\bibitem[{Camera Calibration and 3D Reconstruction}(Accessed: Dec 5,
  2019)]{homography_opencv}
{Camera Calibration and 3D Reconstruction}.
\newblock
  Available online: \url{https://docs.opencv.org/2.4/modules/calib3d/doc/camera\_calibration\_and\_3d\_reconstruction.html} (accessed on 5 December 2019).

\bibitem[{Geometric Image Transformations}(Accessed: Dec 5, 2019)]{warp_opencv}
{Geometric Image Transformations}.
\newblock
  Available online: \url{https://docs.opencv.org/2.4/modules/imgproc/doc/geometric\_transformations.html} (accessed on 5 December 2019).

\bibitem[Rai and Khanna(2012)]{rai2012gender}
Rai, P.; Khanna, P.
\newblock Gender classification techniques: A review. In {\em Advances in
  Computer Science, Engineering \& Applications}; Springer:  Berlin/Heidelberg, Germany, 
 2012; pp. 51--59.

\bibitem[Assael \em{et~al.}(2016)Assael, Shillingford, Whiteson, and
  de~Freitas]{DBLP:journals/corr/AssaelSWF16}
Assael, Y.M.; Shillingford, B.; Whiteson, S.; de~Freitas, N.
\newblock LipNet: Sentence-level Lipreading.
\newblock {\em arXiv} {\bf 2016}, arXiv:1611.01599.

\bibitem[Zeiler(2012)]{zeiler2012adadelta}
Zeiler, M.D.
\newblock Adadelta: an adaptive learning rate method.
\newblock {\em arXiv} {\bf 2012}, arXiv:1212.5701.

\bibitem[Bahdanau \em{et~al.}(2015)Bahdanau, Cho, and
  Bengio]{DBLP:journals/corr/BahdanauCB14}
Bahdanau, D.; Cho, K.; Bengio, Y.
\newblock Neural Machine Translation by Jointly Learning to Align and
  Translate.
\newblock   In \emph{Conference Track Proceedings, Proceedings of the 3rd International Conference on Learning Representations, {ICLR}
  2015, San Diego, CA, USA, 7--9 May 2015};
  Bengio, Y., LeCun, Y., Eds.; DBLP: Trier, Germany, 2015.

\bibitem[Goodfellow \em{et~al.}(2014)Goodfellow, Pouget-Abadie, Mirza, Xu,
  Warde-Farley, Ozair, Courville, and Bengio]{NIPS2014_5423}
Goodfellow, I.; Pouget-Abadie, J.; Mirza, M.; Xu, B.; Warde-Farley, D.; Ozair,
  S.; Courville, A.; Bengio, Y.
\newblock Generative Adversarial Nets. In {\em Advances in Neural Information
  Processing Systems}; Ghahramani, Z., Welling, M., Cortes, C., Lawrence, N.D.,
  Weinberger,~K.Q., Eds.; Curran Associates, Inc.: Red Hook, NY, USA, 2014; pp. 2672--2680.

\bibitem[Isola \em{et~al.}(2017)Isola, Zhu, Zhou, and Efros]{isola2017image}
Isola, P.; Zhu, J.Y.; Zhou, T.; Efros, A.A.
\newblock Image-to-image translation with conditional adversarial networks.
\newblock  In Proceedings of the IEEE Conference on Computer Vision and Pattern
  Recognition (CVPR), Honolulu, HI, USA, 21--26 July 2017; pp. 1125--1134.

\bibitem[Zhu \em{et~al.}(2017)Zhu, Park, Isola, and Efros]{zhu2017unpaired}
Zhu, J.Y.; Park, T.; Isola, P.; Efros, A.A.
\newblock Unpaired image-to-image translation using cycle-consistent
  adversarial networks.
\newblock  In Proceedings of the IEEE International Conference on Computer Vision, Venice, Italy, 22--29 October
  2017; pp. 2223--2232.

\bibitem[Park \em{et~al.}(2020)Park, Efros, Zhang, and
  Zhu]{park2020contrastive}
Park, T.; Efros, A.A.; Zhang, R.; Zhu, J.Y.
\newblock Contrastive learning for unpaired image-to-image translation.
\newblock In \emph{European Conference on Computer Vision, Proceedings of the 16th European Conference, Glasgow, UK, 23--28 August 2020}; Springer:  Cham, Switzerland, 
 2020; pp.
  319--345.

\bibitem[Zhang \em{et~al.}(2018)Zhang, Wiliem, Yang, and Lovell]{zhang2018tv}
Zhang, T.; Wiliem, A.; Yang, S.; Lovell, B.
\newblock {TV-GAN}: Generative adversarial network based thermal to visible
  face recognition.
\newblock  In Proceedings of the International Conference on Biometrics (ICB), Gold Coast, Australia, 20--23 February 2018;
  pp. 174--181.

\bibitem[Wang \em{et~al.}(2018)Wang, Chen, and Wu]{wang2018thermal}
Wang, Z.; Chen, Z.; Wu, F.
\newblock Thermal to visible facial image translation using generative
  adversarial networks.
\newblock {\em IEEE Signal Process. Lett.} {\bf 2018}, {\em
  25},~1161--1165.

\bibitem[{Face Recognition}(Accessed: Dec 5, 2019)]{github_face_detector}
{Face Recognition}.
\newblock
  Available online: \url{https://github.com/opencv/opencv/tree/master/samples/dnn/face_detector} (accessed on 5 December 2019).


\bibitem[Heusel \em{et~al.}(2017)Heusel, Ramsauer, Unterthiner, Nessler, and
  Hochreiter]{heusel2017gans}
Heusel, M.; Ramsauer, H.; Unterthiner, T.; Nessler, B.; Hochreiter, S.
\newblock Gans trained by a two time-scale update rule converge to a local nash
  equilibrium.
\newblock {\em arXiv} {\bf 2017}, arXiv:1706.08500.

\bibitem[{Face Recognition}(Accessed: Dec 5, 2019)]{github_face_recognition}
{Face Recognition}.
\newblock Available online: \url{https://github.com/ageitgey/face_recognition} (accessed on 5 December 2019).

\bibitem[Shon and Glass(2020)]{shon2020multimodal}
Shon, S.; Glass, J.
\newblock Multimodal Association for Speaker Verification.
\newblock In Proceedings of the Annual Conference of International Speech
  Communication Association (INTERSPEECH), Online, 25--29 October 2020; pp. 2247--2251.

\end{thebibliography}
\end{document}